\documentclass[sigconf]{acmart}
\AtBeginDocument{%
  }

\setcopyright{acmlicensed}
\copyrightyear{2026}
\acmYear{2026}
\acmDOI{XXXXXXX.XXXXXXX}
\acmConference[Conference acronym 'XX]{Make sure to enter the correct
  conference title from your rights confirmation email}{June 03--05,
  2018}{Woodstock, NY}
\acmISBN{978-1-4503-XXXX-X/2026/06}

\usepackage{makecell}
\usepackage{multirow}
\usepackage{graphicx}
\usepackage{subfig}

\usepackage{amsmath}

\usepackage{amssymb}

\usepackage{booktabs}
\usepackage{array}
\usepackage{tabularx}
\usepackage{algorithm}
\usepackage{algpseudocode}
\usepackage{enumitem}
\usepackage{etoolbox}

\makeatletter
\g@addto@macro\normalsize{%
  \setlength{\abovedisplayskip}{3pt plus 1pt minus 1pt}%
  \setlength{\belowdisplayskip}{3pt plus 1pt minus 1pt}%
  \setlength{\abovedisplayshortskip}{2pt plus 1pt minus 1pt}%
  \setlength{\belowdisplayshortskip}{2pt plus 1pt minus 1pt}%
  \setlength{\jot}{2pt}%
}
\makeatother

\def\tabstd#1{_{\scriptscriptstyle(#1)}}
\def\best#1#2{\textbf{#1}\({}_{\scriptscriptstyle(#2)}\)}
\def\sig#1#2#3{#1\({}^{#3}_{\scriptscriptstyle(#2)}\)}
\def\bestsig#1#2#3{\textbf{#1}\({}^{#3}_{\scriptscriptstyle(#2)}\)}

\newcommand{\compactdisplaymath}{%
  \setlength{\abovedisplayskip}{5pt plus 1pt minus 1pt}%
  \setlength{\belowdisplayskip}{5pt plus 1pt minus 1pt}%
  \setlength{\abovedisplayshortskip}{3pt plus 1pt minus 1pt}%
  \setlength{\belowdisplayshortskip}{4pt plus 1pt minus 1pt}%
  \setlength{\jot}{2pt}%
}

\BeforeBeginEnvironment{equation}{\begingroup\compactdisplaymath}
\AfterEndEnvironment{equation}{\endgroup}

\begin{document}

\title{Beyond Exposure: Optimizing Ranking Fairness with Non-linear Time-Income Functions}
\author{Xuancheng Li}
\email{lixuancheng23@mails.tsinghua.edu.cn}
\authornotemark[1]
\affiliation{%
  \institution{DCST, Tsinghua University}
   \city{Beijing}
  \country{China}
}
\author{Tao Yang}
\email{taoyang@cs.utah.edu}
\affiliation{%
  \institution{Kahlert School of Computing, University of Utah}
  \city{Salt Lake City}
  \state{Utah}
  \country{USA}
}

\author{Yujia Zhou}
\email{zhouyujia@mail.tsinghua.edu.cn}
\affiliation{%
  \institution{DCST, Tsinghua University}
     \city{Beijing}
  \country{China}
}

\author{Qingyao Ai}
\email{aiqingyao@gmail.com}
\affiliation{%
  \institution{DCST, Tsinghua University}
     \city{Beijing}
  \country{China}
}

\author{Yiqun Liu}
\email{yiqunliu@tsinghua.edu.cn}
\affiliation{%
  \institution{DCST, Tsinghua University}
     \city{Beijing}
  \country{China}
}

\renewcommand{\shortauthors}{Trovato et al.}

\begin{abstract}
Ranking is central to information distribution in web search and
recommendation, where systems must balance relevance-based effectiveness with
fairness for item providers. A prominent formulation is \textit{Exposure
Fairness}, which allocates position-induced attention according to item merit;
however, exposure is often only an intermediate allocation signal, since
provider utility may depend on context-dependent exposure-to-income conversion.
To address this limitation, we study fair ranking with context-dependent
provider utility, which we refer to as \textit{income}. We formalize
\textit{Income Fairness}, requiring cumulative provider income to be
proportional to relevance, and develop a corresponding metric with a
context-dependent unit-income function.

We propose \textbf{DIDRF}, a
\textbf{D}ynamic-\textbf{I}ncome-\textbf{D}erivative-aware
\textbf{R}anking \textbf{F}airness algorithm for income-fair ranking. DIDRF
exploits the quadratic structure of income-fairness violations to derive a
prefix-aware and state-calibrated scoring rule that jointly balances
effectiveness and income fairness. Experiments on standard learning-to-rank
datasets with log-calibrated semi-synthetic income environments constructed
from real advertising and e-commerce logs show that DIDRF improves income
fairness over representative fair-ranking baselines while maintaining
competitive ranking effectiveness.
\end{abstract}




% \received{}
% \received[revised]{}
% \received[accepted]{}

\maketitle
\section{Introduction}
\label{INTRODUCTION}

Ranking serves as the core of information distribution in modern information retrieval (IR) systems such as web search engines and recommendation systems. Depending on how information is ranked and presented to users, ranking systems can control how much utility a user or a provider could get from an IR service. Previous studies on ranking optimization usually focus on improving the effectiveness of ranking systems through maximizing user utility \cite{joachims2017accurately,wang2018position}, i.e., putting the items with the highest relevance to user’s need at positions that are most likely to be noticed by users. 
Nowadays, as more and more people recognize that monotonically optimizing relevance could lead to severe imbalance and unfairness to information providers in the market, more research starts to study fairness problems in ranking algorithms.
How to develop a ranking system that can provide relevant results to users while maintaining a reasonable fairness to information providers has thus become an important research question in the IR community \cite{gorantla2024optimizing,xu2024taxation,ye2024bankfair,xu2025elasticity}.  

When talking about ranking fairness for information providers, one of the most well-known concepts is \textit{Exposure Fairness} \cite{mehrabi2021survey}. 
Assuming that information providers could benefit only from the exposure of their products to users, the problem of ranking fairness from the provider side is essentially a problem of exposure allocations.
Therefore, the idea of exposure fairness is that, if a ranking system could provide equal exposure to information providers with similar properties, then the system should be considered fair. 
Representative work in this direction spans group-based fairness
\cite{tsang2019group,kearns2019empirical}, relevance-based fairness
\cite{singh2018fairness}, adaptive exposure allocation
\cite{jaenich2024fairness}, distributed amortized exposure
\cite{xu2024fairsync}, and scalable exposure control
\cite{togashi2024scalable}. 
While the objectives and definitions of fairness could vary a bit, most of the existing studies on ranking fairness focus on exposure fairness and assume that provider utility can be directly represented or is proportional to item exposure \cite{biega2018equity,diaz2020evaluating,brantley2024ranking,balagopalan2025query}. 

Despite its simplicity and popularity, exposure fairness relies on a proxy assumption: position-induced exposure can represent provider-side utility. 
This can be unreliable in practice \cite{patro2022fair} because provider
utility is not determined by exposure alone: beyond exposure, it often depends
on context-dependent factors such as time and region, which affect how exposure
is converted into downstream outcomes. Thus, rankings allocate exposure
opportunities, but high placements may not necessarily translate into
comparable realized provider utility. Even for items with comparable relevance,
the value realized from the same exposure may vary across temporal, spatial, or
other contexts.
Time-sensitive domains provide intuitive examples. For example, early exposure is more valuable in breaking news
\cite{chakraborty2017optimizing}, while restaurant utility may vary with both
time and location, such as whether nearby customers are active during peak hours
\cite{yuan2013time,banerjee2020analyzing}. Thus, exposure is only an
intermediate allocation signal, and equal exposure may still yield unequal
realized provider utility.
% For instance, time also represents a pivotal factor affecting the provider utility \cite{patro2022fair}. In some time-sensitive domains, the utility realized by providers is time-dependent: earlier exposure in breaking news consistently translates into higher news provider income \cite{chakraborty2017optimizing} and restaurants obtain more orders when recommended to nearby customers during peak hours \cite{yuan2013time, banerjee2020analyzing}.
% Therefore, exposure should be viewed as an intermediate allocation signal: equalizing exposure may still leave realized provider-side utility unequal.

Therefore, this paper studies fair ranking when exposure does not fully capture the final provider-side utility. 
We define the provider utility induced by exposure in ranking as \textbf{income}, i.e., a ranking-induced expected provider-utility signal under a given context (e.g., time). We establish a formal definition of \textit{Income Fairness}, introduce an income function that maps context-specific exposure to income, and develop a corresponding measurement metric. Instead of modeling provider utility solely using item exposure, income fairness seeks to ensure fairness in provider utility within ranking under more realistic, complex, and context-dependent conditions.

Optimizing income fairness is challenging because the ranking system directly
controls exposure rather than income. Changing an item's ranking position
changes its allocated exposure, but the income produced by one unit of exposure
may vary with context. Among various contextual factors, time is a
representative one, as temporal signals have been widely used to model
context-dependent behavior and outcomes \cite{campos2014time,yuan2013time}.
Therefore, in our experiments, we instantiate this idea through time-varying
unit income to capture temporal changes in provider utility. We construct two
log-calibrated semi-synthetic environments to evaluate whether exposure-based
fair-ranking methods remain fair when the mapping from exposure to provider
utility is context-dependent. In our evaluation, we adapt representative
exposure fairness algorithms to the proposed income fairness objective.
However, these adaptations do not fully model the dynamic coupling between
ranking decisions, assigned exposure, and context-dependent unit income, and
therefore may be less effective at optimizing income fairness in this setting.

To address this limitation, we propose the \textbf{D}ynamic-\textbf{I}ncome-\textbf{D}erivative-aware \textbf{R}anking \textbf{F}airness (DIDRF) algorithm for both offline and online ranking settings. 
DIDRF jointly optimizes effectiveness and income fairness by estimating the marginal effect of each ranking decision on accumulated provider income. 
Rather than simply replacing exposure with income, DIDRF derives a marginal objective that explicitly models how the current ranking-induced income increment changes cumulative income fairness.  
This yields a prefix-aware and state-calibrated ranking score that combines effectiveness, an income-deficit signal, and a second-order fairness correction. 
Experiments show that DIDRF provides a better effectiveness--fairness trade-off than income-adapted variants of representative exposure-fairness algorithms. \footnote{Our code is available at https://anonymous.4open.science/r/DIDRF}
\vspace{-1.0em}

\section{RELATED WORK}

\textbf{Exposure Fairness and Income Fairness.}
Exposure fairness is a central notion in fair ranking: Biega et al. \cite{biega2018equity} proposed equity of attention, and Singh and Joachims \cite{singh2018fairness} formalized fairness of exposure in rankings. 
These methods mitigate position bias by requiring accumulated exposure to be proportional to relevance or merit \cite{craswell2008experimental}, and recent work extends this view to adaptive allocation, scalable recommendation, long-term constraints, and query-aware amortized ranking \cite{jaenich2024fairness,xu2024fairsync,togashi2024scalable,brantley2024ranking,balagopalan2025query}. 
However, fair exposure does not necessarily imply fair income, since income can
also depend on contextual factors, as discussed in \S\ref{INTRODUCTION}.
Income fairness has been studied more broadly in distributive justice and algorithmic allocation \cite{alves1978should,black2022algorithmic}, but remains underexplored in fair ranking, where ranking decisions allocate exposure while realized income may vary across ranking contexts.

\textbf{Fair Ranking Algorithms.}
Fair ranking algorithms are often categorized into open-loop and feedback-loop
methods based on whether historical outcomes are used to update ranking scores
\cite{yang2023fara}. Open-loop methods optimize rankings under fixed relevance
estimates using linear programming, policy-gradient optimization, differentiable
ranking models, or fairness-aware learning-to-rank objectives
\cite{heuss2022fairness,singh2018fairness,singh2019policy,oosterhuis2021computationally,gorantla2024optimizing}.
Feedback-loop methods update ranking policies using accumulated exposure,
fairness deficits, or long-term ranking constraints
\cite{morik2020controlling,yang2021maximizing,yang2023marginal,yang2023fara},
while recent variants further consider economic or traffic-aware allocation
\cite{xu2024taxation,ye2024bankfair,xu2025elasticity}. Different from these exposure-centered methods, DIDRF is designed for the more
challenging income-fairness setting and provides a dedicated optimization
framework for this objective.

\begin{figure}[t]
    \centering
    \includegraphics[width=0.48\textwidth]{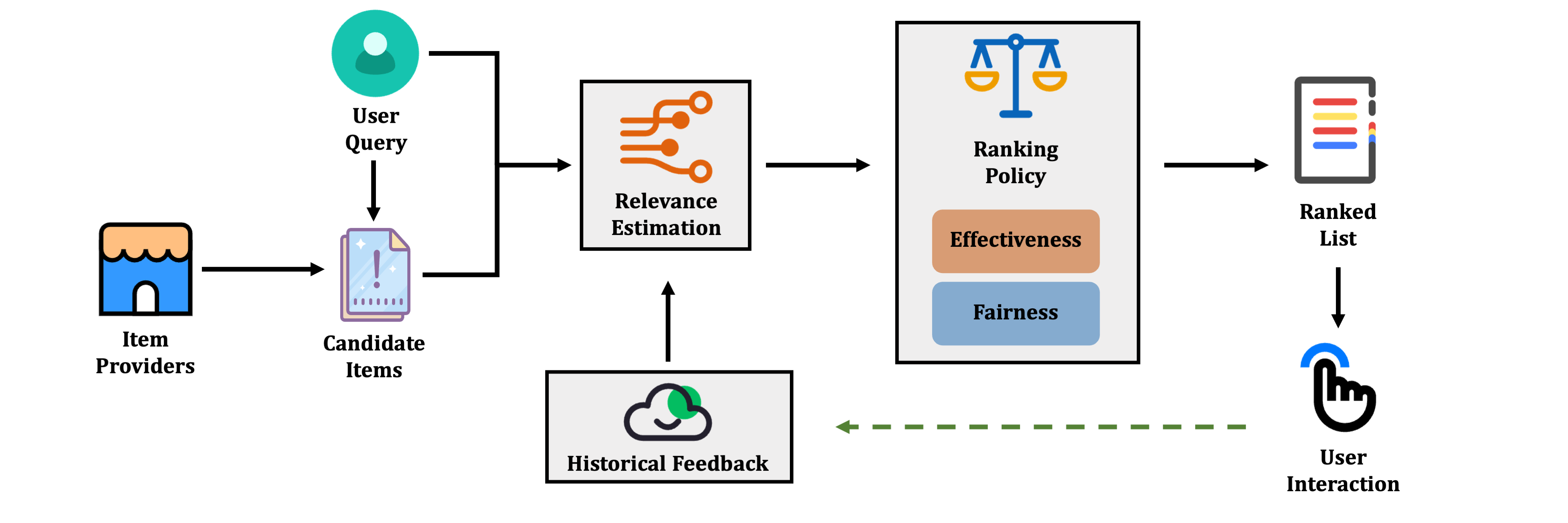}
    \caption{Workflow of the dynamic ranking system.}
    \label{fig:workflow}
    \vspace{-1.5em}
\end{figure}
\section{Background}
Ranking systems operate as two-sided mechanisms: they serve users by placing
relevant items at positions likely to be examined, while allocating exposure
opportunities to item providers. In dynamic settings, the displayed rankings
also shape biased feedback signals, which in turn affect future ranking
decisions and downstream outcomes such as income. We therefore first
formalize the dynamic ranking workflow and feedback model used throughout the
paper, and then review the effectiveness objective and
exposure-based fairness metrics built on this workflow. Finally, we
explain why exposure alone can be an insufficient proxy for provider utility,
motivating the income-based fairness formulation introduced next.

\subsection{The Ranking System}
\label{3.1}

\subsubsection{Ranking Workflow}

Figure~\ref{fig:workflow} illustrates the dynamic ranking workflow considered in this paper. 
Building on prior research \cite{yang2023marginal,yang2023fara}, the ranking process is conventionally structured as follows, and our study follows the same general procedure. 
At timestep \(i\), a ranking session is initiated when a user submits a query \(q\).\footnote{The query \(q\) represents a general ranking query, such as a recommendation query or a text query submitted to a search engine.} 
For this query, the system retrieves a candidate set \(D(q)\) from item providers. 
If the true relevance of these items is unknown, the system uses a relevance estimator to infer item relevance from historical feedback. 
The system then constructs a ranked list \(\pi_i\), where \(\pi_i[j]\) denotes the item placed at position \(j\), by optimizing ranking objectives such as effectiveness, fairness, or a combination of both. 
After the ranked list is presented to the user, feedback signals such as clicks are collected and used to update the relevance estimator. 
Meanwhile, the system maintains accumulated ranking outcomes, such as exposure or income, which can be used to support subsequent optimization of the ranking policy.

\subsubsection{Partial and Biased Feedback}

As noted above, user feedback plays a critical role in ranking systems. 
However, it can be a partial and biased indicator of relevance, because users can only provide meaningful feedback for items that they have examined. 
Following existing works on click models \cite{chuklin2022click,yang2021maximizing}, we model the probability of a user clicking on item \(d\) in ranking list \(\pi_i\) for query \(q\) as:
\begin{equation}
\small
p(c=1\mid d,\pi_i,q)
=
p(e=1\mid d,\pi_i)\,p(r=1\mid d,q),
\label{eq:1}
\end{equation}
where \(p(e=1\mid d,\pi_i)\) is the probability that item \(d\) is examined by the user, and \(p(r=1\mid d,q)\) is the probability that item \(d\) is relevant to query \(q\). 
This factorization indicates that a click is observed only when the item is both examined and relevant. 
Therefore, clicks do not directly reveal relevance: an unclicked item may be irrelevant, or it may simply not have been examined. 
This motivates us to explicitly model examination bias in the ranking process.

Following existing works \cite{oosterhuis2021computationally,yang2022can}, we consider two types of bias in the user's examination process. 
First, \textbf{positional bias} means that the examination probability depends on the item's position in the ranked list and usually decreases as the position moves down the list \cite{craswell2008experimental}. 
Second, \textbf{selection bias} occurs when not all items are presented to users or when the ranked list is too long for users to examine completely \cite{oosterhuis2020policy}. 
We capture these two biases using a position-dependent examination probability with an examination cutoff \(k_c\):
\begin{equation}
\small
p(e=1\mid d,\pi_i)
=
\begin{cases}
p_{\operatorname{rank}(d\mid\pi_i)}, 
& \text{if } \operatorname{rank}(d\mid\pi_i)\le k_c,\\
0, 
& \text{otherwise},
\end{cases}
\label{eq:2}
\end{equation}
where \(\operatorname{rank}(d\mid\pi_i)\) denotes the position of item \(d\) in \(\pi_i\), and \(p_j\) denotes the examination probability at position \(j\). 

\subsection{User Utility Measurement}
\label{3.2}

In ranking systems, user satisfaction, also referred to as user-side utility or effectiveness, is crucial for evaluating performance. 
It measures whether relevant items are placed at positions that are likely to be examined. 
A widely used metric for user-side utility is \textbf{Normalized Discounted Cumulative Gain (NDCG)} \cite{jarvelin2002cumulated}. 
For a ranking list \(\pi_i\) displayed for query \(q\), we use:
\begin{equation}
\small
\mathrm{NDCG}@k_{c}(\pi)
= \frac{\sum_{i=1}^{k_{c}} R(\pi[i],q)\,\lambda_i}{\sum_{i=1}^{k_{c}} R(\pi^{*}[i],q)\,\lambda_i}
= \frac{\sum_{i=1}^{k_{c}} R(\pi[i],q)\,p_i}{\sum_{i=1}^{k_{c}} R(\pi^{*}[i],q)\,p_i}.
\label{eq:ndcg}
\end{equation}
where \(R(\pi_i[j],q)\)\footnote{Throughout the rest of the paper, when the fixed query \(q\) is clear from context, we omit \(q\) and write the relevance score as \(R(\pi_i[j])\) or \(R(d)\).} is the relevance score of the item at position \(j\), \(\pi_i^*\) is the ideal ranking sorted by relevance, and $\lambda_i$ is the weight assigned to position $i$. Following prior work \cite{saito2022fair}, we set $\lambda_i$ equal to the examination probability $p_i$. 

Building on NDCG, \textbf{cumulative NDCG (cNDCG)} is commonly used to evaluate effectiveness in online ranking systems \cite{yang2023marginal,yang2023fara}:
\begin{equation}
\small
\mathrm{Eff}(n)
=
\mathrm{cNDCG}@k_c(n)
=
\sum_{i=1}^{n}
\alpha^{n-i}
\mathrm{NDCG}@k_c(\pi_i,q),
\label{eq:cndcg}
\end{equation}
where \(\alpha\in(0,1]\) is a temporal discount factor and \(n\) is the total number of ranking timesteps. We also use cNDCG as the evaluation metric for effectiveness in this paper.

\subsection{Provider Utility Measurement}
\label{3.3}

The order in which items are ranked by a ranking system can have a significant impact on the income generated by item providers. 
A fair ranking system should ensure that items with similar quality, i.e., relevance, receive comparable treatment, ultimately aiming to provide item providers with similar earnings for items of similar relevance. 
In addition to ensuring the effectiveness of the ranking system from the user's perspective, it is also important to consider fairness towards items and item providers from the provider's perspective.

The most commonly employed fairness metric in current online ranking algorithms is \textbf{Exposure Fairness}. 
Exposure fairness treats exposure, defined via cumulative examination probabilities, as a proxy for provider utility. 
Intuitively, exposure fairness requires that the cumulative exposure received by each document be proportional to its relevance, so that more relevant documents receive correspondingly greater exposure over time.
Following the standard cumulative exposure formulation in prior work \cite{biega2018equity}, the cumulative exposure of item \(d\) after \(n\) ranking timesteps is given by:
\begin{equation}
\small
E(d,n)
=
\sum_{i=1}^{n}
\sum_{j=1}^{k_c}
p_j\,\mathbb{I}\!\left(\pi_i[j]=d\right).
\label{eq:cumulative_exposure}
\end{equation}
where \(E(d,n)\) denotes the cumulative exposure received by item \(d\) over \(n\) ranking timesteps, and \(\mathbb{I}(\cdot)\) is the indicator function.
Based on cumulative exposure, we quantify exposure unfairness using the
following metric \cite{oosterhuis2021computationally}:
\begin{equation}
\small
\begin{aligned}
\mathrm{unfair}_{\mathrm{exp}}(q,n)
&=
\frac{1}{|D(q)|(|D(q)|-1)}
\\[-0.3ex]
&\quad
\sum_{\substack{d_x,d_y\in D(q)\\ d_x\ne d_y}}
\Big(
E(d_x,n)R(d_y)
-
E(d_y,n)R(d_x)
\Big)^2,
\\
\mathrm{fair}_{\mathrm{exp}}(q,n)
&=
-\mathrm{unfair}_{\mathrm{exp}}(q,n).
\end{aligned}
\label{eq:exposure_unfairness}
\end{equation}
where \(D(q)\) is the candidate item set for query \(q\). 
Exposure fairness provides a clear and widely used way to model fairness by treating exposure as the opportunity allocated to item providers \cite{biega2018equity,singh2018fairness,oosterhuis2021computationally}. 
However, exposure is only a proxy for the income that providers eventually realize. 
When the mapping from exposure to income varies across items or contexts, equal exposure need not imply equal income. 
This motivates the income fairness formulation introduced in \S\ref{sec:income_fairness}.
\section{Income Fairness}
\label{sec:income_fairness}

Exposure fairness captures the opportunity allocated by a ranking, but the
income realized from that opportunity may vary. We use \textit{income} broadly
to denote measurable provider utility induced by ranking exposure, including
revenue, orders, conversions, and other benefits.
The gap between exposure and income becomes especially clear when the value of
exposure depends on context. For example, in news platforms, earlier exposure
can generate substantially more income than later exposure, especially for newly
published stories \cite{chakraborty2017optimizing}. In local restaurant
ranking, showing a restaurant to nearby users during peak hours may bring more
income than the same exposure delivered off-peak
\cite{yuan2013time,banerjee2020analyzing}. These examples show that the
exposure-to-income mapping can depend on temporal, spatial, or market contexts. Therefore, a ranking that satisfies exposure
fairness may still produce unequal realized income. We thus formalize
\textit{Income Fairness}, which evaluates fairness based on income realized
from ranking exposure rather than exposure alone.
\subsection{Definition of Income Fairness}
\label{sec:income_fairness_definition}

Following the fixed-query setting in \S\ref{3.3}, we define income fairness over the candidate set \(D(q)\). 
Let \(v_d^i\) denote the ranking-induced income of item
\(d\) at timestep \(i\), i.e., the measurable provider gains generated
from exposure at that timestep.

\begin{definition}[Income Fairness]
A ranking at timestep \(i\) satisfies \textit{Income Fairness} if the income realized by each item is proportional to its relevance. 
Equivalently, for any two items \(d_x,d_y\in D(q)\) with positive relevance scores, their income-to-relevance ratios are equal:
\begin{equation}
\small
\frac{v_{d_x}^i}{R(d_x)}
=
\frac{v_{d_y}^i}{R(d_y)}.
\label{eq:income_fairness}
\end{equation}
\end{definition}

\textit{Income Fairness} may not hold exactly at each individual ranking timestep, especially when similarly relevant items compete for limited exposure. 
Following the idea of amortized fairness in ranking \cite{biega2018equity,singh2018fairness}, we therefore define income fairness over cumulative income across multiple ranking timesteps. 
For item \(d\), its cumulative income after \(n\) ranking timesteps is:
\begin{equation}
\small
I(d,n)
=
\sum_{i=1}^{n}
v_d^i.
\label{eq:cumulative_income_general}
\end{equation}

\begin{definition}[Amortized Income Fairness]
For a fixed query \(q\), a sequence of rankings \(\pi_1,\pi_2,\dots,\pi_n\) satisfies \textit{Amortized Income Fairness} if each item's cumulative income is proportional to its relevance. 
Equivalently, for any two items \(d_x,d_y\in D(q)\) with positive relevance scores, their cumulative income-to-relevance ratios are equal:
\begin{equation}
\small
\frac{I(d_x,n)}{R(d_x)}
=
\frac{I(d_y,n)}{R(d_y)}.
\label{eq:amortized_income_fairness}
\end{equation}
\end{definition}

\subsection{Income Measurement}
\label{sec:income_measurement}

In this section, we introduce a measure of income fairness for general ranking systems.
Our starting point is that an item can generate income for its provider only when it is examined by the user. 
Moreover, conditional on being examined, the expected income per examination may depend on the ranking context. 
Therefore, the ranking-induced income of an item can be decomposed into two components: 
(i) \emph{exposure}, which determines how likely the item is examined under a displayed ranking, and 
(ii) a \emph{unit-income} term, which quantifies the expected income generated per unit exposure under the current context.

Based on this decomposition, we measure the income of item \(d\) at timestep \(i\) as the product of its position-induced exposure and a context-dependent unit-income function. 
Let \(\xi_i\) denote the ranking context at timestep \(i\), such as temporal, spatial, or market conditions. 
We use \(\mu(d,q,\xi_i)\ge 0\) to denote the \emph{unit-income function}, i.e.,
the expected income generated by one unit of exposure for item \(d\) under
query \(q\) and context \(\xi_i\). In our implementation, we normalize
\(\mu(d,q,\xi_i)\) to \([0,1]\) across timesteps to ensure comparable unit-income
scales. Given the displayed ranking \(\pi_i\), the income associated with item
\(d\) at timestep \(i\) is then defined as:
\begin{equation}
\small
\Delta E(d,i)
=
\sum_{j=1}^{k_c}
p_j\,\mathbb{I}\!\left(\pi_i[j]=d\right).
\label{eq:marginal_exposure_income}
\end{equation}
Here, \(\Delta E(d,i)\) measures the expected exposure of item \(d\) at timestep \(i\), or equivalently, how likely item \(d\) is to be examined under the displayed ranking \(\pi_i\). The income earned by item \(d\) at timestep \(i\) is then defined as:
\begin{equation}
\small
v_d^i
=
\Delta I(d,i)
=
\Delta E(d,i)\ \mu(d,q,\xi_i)
=
\left(
\sum_{j=1}^{k_c}
p_j\,\mathbb{I}\!\left(\pi_i[j]=d\right)
\right)
\mu(d,q,\xi_i).
\label{eq:marginal_income}
\end{equation}

This formulation cleanly separates the position-driven exposure component \(\Delta E(d,i)\), which is controlled by the ranking system through examination probabilities, from the context-dependent utility-conversion component \(\mu(d,q,\xi_i)\), which captures how each unit of exposure is converted into income.
Over \(n\) timesteps, the cumulative income of item \(d\) is:
\begin{equation}
\small
I(d,n)
=
\sum_{i=1}^{n}
v_d^i
=
\sum_{i=1}^{n}
\left[
\left(
\sum_{j=1}^{k_c}
p_j\,\mathbb{I}\!\left(\pi_i[j]=d\right)
\right)
\mu(d,q,\xi_i)
\right].
\label{eq:cumulative_income}
\end{equation}

In this paper, we mainly instantiate \(\xi_i\) with the time factor to model
time-varying unit income, while the formulation can incorporate other
contextual variables when available.

\subsection{Income Fairness Metric}
\label{sec:income_fairness_metric}

Following Eq.~\eqref{eq:amortized_income_fairness}, we measure income
unfairness as the average squared pairwise violation of income--relevance
proportionality. For a fixed query \(q\), let \(m=|D(q)|\). After \(n\)
ranking timesteps, the query-level income unfairness is:
\begin{equation}
\small
\begin{aligned}
\mathrm{unfair}_{\mathrm{inc}}(q,n)
&=
\frac{1}{m(m-1)}
\sum_{\substack{d_x,d_y\in D(q)\\ d_x\ne d_y}}
\Big(
I(d_x,n)R(d_y)
-
I(d_y,n)R(d_x)
\Big)^2,
\\
\mathrm{fair}_{\mathrm{inc}}(q,n)
&=
-\mathrm{unfair}_{\mathrm{inc}}(q,n).
\end{aligned}
\label{eq:income_unfairness}
\end{equation}

For efficient computation, Eq.~\eqref{eq:income_unfairness} can be rewritten
without explicitly enumerating all item pairs:
\begin{equation} \small \begin{aligned} \mathrm{unfair}_{\mathrm{inc}}(q,n) &= \frac{2}{m(m-1)} [ (\sum_{d\in D(q)}R(d)^2) (\sum_{d\in D(q)}I(d,n)^2) \\[-0.3ex] &\qquad\qquad - (\sum_{d\in D(q)}I(d,n)R(d))^2]. \end{aligned} \label{eq:income_unfairness_scalar_sum} \end{equation}
The equality follows by expanding the ordered pairwise sum in
Eq.~\eqref{eq:income_unfairness}. Thus, the metric can be computed using only
three scalar sums over \(D(q)\).
For a query set \(\mathcal{Q}\), we report the average query-level unfairness:
\begin{equation}
\small
\mathrm{unfair}_{\mathrm{inc}}(n)
=
\frac{1}{|\mathcal{Q}|}
\sum_{q\in\mathcal{Q}}
\mathrm{unfair}_{\mathrm{inc}}(q,n),
\qquad
\mathrm{fair}_{\mathrm{inc}}(n)
=
-\mathrm{unfair}_{\mathrm{inc}}(n).
\label{eq:overall_income_unfairness}
\end{equation}

\section{Method}
\label{sec:method}

% We develop DIDRF, a \textbf{D}ynamic-\textbf{I}ncome-\textbf{D}erivative-aware
% \textbf{R}anking \textbf{F}airness algorithm for income-fair ranking.
% DIDRF jointly optimizes ranking effectiveness and income fairness in a dynamic
% ranking process. We formulate a one-step marginal objective for the current
% ranking timestep and decompose it into effectiveness and income-fairness terms.
% By differentiating cumulative income unfairness with respect to ranking-induced
% income, DIDRF derives an income-deficit signal and a second-order marginal
% fairness gain. These terms are then combined into a prefix-aware and
% state-calibrated ranking score, which we instantiate in both offline and online
% settings before analyzing its time complexity.

We develop DIDRF, a \textbf{D}ynamic-\textbf{I}ncome-\textbf{D}erivative-aware
\textbf{R}anking \textbf{F}airness algorithm for income-fair ranking. DIDRF
jointly optimizes ranking effectiveness and income fairness through a one-step
marginal objective in a dynamic ranking process. By differentiating cumulative
income unfairness with respect to ranking-induced income, DIDRF obtains an
income-deficit signal and a second-order marginal fairness correction. These
signals are combined into a prefix-aware and state-calibrated ranking score,
which is instantiated in both offline and online settings.

DIDRF focuses on fair ranking given an available income signal, rather than on
income simulation or income estimation itself. In other words, our goal is to
optimize ranking decisions when exposure induces heterogeneous,
context-dependent provider income across items. Income modeling and simulation are orthogonal to our ranking objective and can
be supplied by external estimation or microsimulation modules when needed
~\cite{gasior2024assessing,toder2024use,bronka2025simpaths};
accordingly, the following derivations assume that the unit-income signal
\(\mu(d,q,\xi_n)\) is available.

\subsection{Objective}
\label{sec:didrf_objective_v2}

At the \(n\)-th ranking timestep for query \(q\), the system retrieves a
candidate set \(D(q)\). Let \(m=|D(q)|\) denote the number of candidate items.
The system then returns a ranked list \(\pi_n\). Formally, DIDRF considers the
following effectiveness--fairness objective:
\begin{equation}
\small
\pi_n^\star
=
\arg\max_{\pi_n}
\mathrm{Obj}(q,n),
\qquad
\mathrm{Obj}(q,n)
=
\mathrm{Eff}(q,n)
+
\gamma \mathrm{fair}_{\mathrm{inc}}(q,n),
\label{eq:didrf_full_objective_v2}
\end{equation}
where \(\gamma\ge0\) controls the tradeoff between user effectiveness
defined in Eq.~\eqref{eq:cndcg} and income fairness derived from
Eq.~\eqref{eq:income_unfairness}. Following prior marginal and feedback-based
fair-ranking methods
\cite{yang2023marginal,yang2021maximizing,gao2022fair}, directly optimizing
the full cumulative objective at each timestep is computationally challenging:
the income-fairness term naively involves \(O(m^2)\) item-pair terms, and
optimizing over ranked lists further induces a combinatorial decision space.

Since the current ranking \(\pi_n\) cannot change the previous objective value
\(\mathrm{Obj}(q,n-1)\), DIDRF instead optimizes the one-step marginal gain:
\begin{equation}
\small
\begin{aligned}
\pi_n^\star
&=
\arg\max_{\pi_n}
\Delta \mathrm{Obj}(q,n),
\\
\Delta \mathrm{Obj}(q,n)
&=
\mathrm{Obj}(q,n)-\mathrm{Obj}(q,n-1)
\\
&=
\Delta \mathrm{Eff}(q,n)
+
\gamma \Delta \mathrm{fair}_{\mathrm{inc}}(q,n).
\end{aligned}
\label{eq:didrf_delta_objective_v2}
\end{equation}
This marginal decomposition isolates the incremental effect of the current
ranking decision, which is the only part of the objective controllable at
timestep \(n\).

\subsection{Marginal Effectiveness Optimization}
\label{sec:didrf_effectiveness_v2}

Following the marginal-optimization perspective in
\cite{yang2023marginal}, we first handle the marginal effectiveness term
\(\Delta \mathrm{Eff}(q,n)\) in Eq.~\eqref{eq:didrf_delta_objective_v2}.
Since the effectiveness component is linear in exposure, its first-order
expansion is exact:
\begin{equation}
\small
\Delta \mathrm{Eff}(q,n)
=
\sum_{d\in D(q)}
\frac{\partial \mathrm{Eff}(q,n)}
{\partial E(d,n)}
\Delta E(d,n)
=
\sum_{d\in D(q)}
R^+(d)\Delta E(d,n).
\label{eq:didrf_delta_eff_v2}
\end{equation}
Here, \(\Delta E(d,n)\) is the marginal exposure assigned to item \(d\) by the
current ranking, as defined in Eq.~\eqref{eq:marginal_exposure_income}.
\(R^+(d)\) is a relevance-based utility signal for the effectiveness objective;
its instantiation depends on whether relevance is known or estimated, as
detailed in \S\ref{sec:didrf_offline_online_v2}. Thus, placing item \(d\) at
rank \(j\) gives \(\Delta E(d,n)=p_j\) and contributes \(p_jR^+(d)\) to the
marginal effectiveness objective. From the effectiveness perspective, items
with larger \(R^+(d)\) should receive larger position exposure.

\subsection{Marginal Income-Fairness Optimization}
\label{sec:didrf_income_fairness_v2}

We next handle the marginal income-fairness term
\(\Delta \mathrm{fair}_{\mathrm{inc}}(q,n)\) in
Eq.~\eqref{eq:didrf_delta_objective_v2}. Unlike effectiveness, income fairness
depends on cumulative income, whereas the current ranking directly controls
exposure. Using the unit-income function defined in
\S\ref{sec:income_measurement}, the exposure assigned by the current ranking
induces the following income increment:
\begin{equation}
\small
\Delta I(d,n)=\mu(d,q,\xi_n)\Delta E(d,n),
\label{eq:didrf_delta_income_v2}
\end{equation}
for each \(d\in D(q)\).

Let \(I^-(d)=I(d,n-1)\) denote the pre-ranking cumulative income. For
notational clarity, write \(\mathrm{fair}_{\mathrm{inc}}(q;I)\) as the
income-fairness objective evaluated at cumulative income vector \(I\). Following
the income-fairness metric in Eq.~\eqref{eq:income_unfairness},
\(\mathrm{fair}_{\mathrm{inc}}(q;I)\) is a quadratic function of \(I\).
Therefore, its second-order expansion from \(I^-\) with respect to the current
income increments is exact:
\begin{equation}
\small
\begin{aligned}
\Delta \mathrm{fair}_{\mathrm{inc}}(q,n)
&=
\mathrm{fair}_{\mathrm{inc}}(q;I^-+\Delta I)
-
\mathrm{fair}_{\mathrm{inc}}(q;I^-)
\\
&=
\sum_{d\in D(q)}
\left.
\frac{\partial \mathrm{fair}_{\mathrm{inc}}(q;I)}
{\partial I(d)}
\right|_{I=I^-}
\Delta I(d,n)
\\
&\quad+
\frac{1}{2}
\sum_{d_x,d_y\in D(q)}
\left.
\frac{\partial^2 \mathrm{fair}_{\mathrm{inc}}(q;I)}
{\partial I(d_x)\partial I(d_y)}
\right|_{I=I^-}
\Delta I(d_x,n)\Delta I(d_y,n).
\end{aligned}
\label{eq:didrf_fairness_taylor_v2}
\end{equation}

Define:
\begin{equation}
\small
\sigma=\frac{4}{m(m-1)}.
\label{eq:didrf_sigma_v2}
\end{equation}
The first-order derivative evaluated at the pre-ranking income state is:
\begin{equation}
\small
\left.
\frac{\partial \mathrm{fair}_{\mathrm{inc}}(q;I)}
{\partial I(d)}
\right|_{I=I^-} = 
\underbrace{\sigma
\left(
R(d)\sum_{e\in D(q)}I^-(e)R(e)
-
I^-(d)\sum_{e\in D(q)}R(e)^2
\right)}_{g(d,n)}.
\label{eq:didrf_gradient_v2}
\end{equation}
The second-order derivative, which is constant in the income state, is:
\begin{equation}
\small
\frac{\partial^2 \mathrm{fair}_{\mathrm{inc}}(q;I)}
{\partial I(d_x)\partial I(d_y)}
=
-\sigma
\left(
\mathbb{I}(d_x=d_y)\sum_{e\in D(q)}R(e)^2
-
R(d_x)R(d_y)
\right).
\label{eq:didrf_hessian_v2}
\end{equation}
The second-order partial derivative in Eq.~\eqref{eq:didrf_hessian_v2}
captures both item-specific curvature and cross-item interaction under the
relevance entitlement \(R(d)\) used by the income-fairness objective. We
distinguish \(R(d)\) from the relevance-based effectiveness signal \(R^+(d)\);
their instantiations under known and estimated relevance are specified in
\S\ref{sec:didrf_offline_online_v2}. Substituting
Eq.~\eqref{eq:didrf_delta_income_v2} into
Eq.~\eqref{eq:didrf_fairness_taylor_v2} turns the marginal income-fairness gain
into an exposure-controlled objective. Thus, the current ranking controls both
marginal objectives through exposure: effectiveness is weighted by \(R^+(d)\),
while income fairness is computed through \(R(d)\) and the exposure-induced
income increment.

% \subsection{DIDRF Score and Ranking Policy}
% \label{sec:didrf_score_policy_v2}

% We now convert the marginal objective derived above into a practical ranking
% policy. At rank \(j\), DIDRF greedily selects the item with the largest
% candidate-dependent marginal score, conditioned on the prefix that has already
% been constructed. For compactness, define:
% \begin{equation}
% \small
% \delta_j(d)=\mu(d,q,\xi_n)p_j,
% \qquad
% A=\sum_{e\in D(q)}R(e)^2 .
% \label{eq:didrf_delta_A_v2}
% \end{equation}
% Here, \(\delta_j(d)\) is the income increment obtained by assigning item \(d\)
% to rank \(j\).

\subsection{DIDRF Score and Ranking Policy}
\label{sec:didrf_score_policy_v2}

We now convert the marginal objective derived above into a practical ranking
policy. The exact maximization of the one-step marginal objective over all
ranked lists remains combinatorial. DIDRF therefore uses a greedy prefix-aware
approximation that selects one item at a time according to the
candidate-dependent marginal score derived below. At rank \(j\), the score is
conditioned on the prefix that has already been constructed. For compactness,
define:
\begin{equation}
\small
\delta_j(d)=\mu(d,q,\xi_n)p_j,
\qquad
A=\sum_{e\in D(q)}R(e)^2 .
\label{eq:didrf_delta_A_v2}
\end{equation}
Here, \(\delta_j(d)\) is the income increment obtained by assigning item \(d\)
to rank \(j\).
Suppose the first \(j-1\) positions have been filled, and let \(S_{j-1}\)
denote the selected set. The marginal effectiveness contribution of placing
item \(d\) at rank \(j\) is:
\begin{equation}
\small
G^{\mathrm{eff}}_j(d)=p_jR^+(d).
\label{eq:didrf_eff_gain_v2}
\end{equation}
For the income-fairness part, the first-order contribution is
\(g(d,n)\delta_j(d)\). The second-order correction is obtained by substituting
Eq.~\eqref{eq:didrf_hessian_v2} into the second-order Taylor term and removing
prefix-only constants that do not affect candidate comparison.\footnote{The
algebraic derivation is provided in \S~\ref{app:fair-gain-derivation}.}
We retain the key candidate-dependent result below. Define the prefix
interaction state as:
\begin{equation}
\small
B_{j-1}
=
\sum_{s<j}
R(\pi_n[s])\delta_s(\pi_n[s])
=
\sum_{s<j}
R(\pi_n[s])\mu(\pi_n[s],q,\xi_n)p_s .
\label{eq:didrf_prefix_state_v2}
\end{equation}
Then, the raw marginal fairness gain of placing item \(d\) at rank \(j\) is:
\begin{equation}
\small
G^{\mathrm{fair}}_j(d)
=
\delta_j(d)
\left[
g(d,n)+\sigma R(d)B_{j-1}
\right]
-
\frac{\sigma}{2}
\left(
A-R(d)^2
\right)
\delta_j(d)^2 .
\label{eq:didrf_fair_gain_v2}
\end{equation}
The term \(g(d,n)\delta_j(d)\) captures the first-order income-deficit signal,
\(\sigma R(d)B_{j-1}\delta_j(d)\) captures the interaction between the current
candidate and the selected prefix, and the last term is the item-level
second-order correction.

The above \(G^{\mathrm{fair}}_j(d)\) is the raw fairness gain induced by the
Taylor approximation. To make the ranking policy adaptive to the current
income state, we further apply a bounded state calibration. Let
\(I^{(j-1)}(d,n)\) be the temporary cumulative income after the first \(j-1\)
positions have been filled, and define the residual income unfairness as:
\begin{equation}
\small
\mathcal{U}_{j-1}
=
\mathrm{unfair}_{\mathrm{inc}}\!\left(q; I^{(j-1)}\right).
\label{eq:didrf_residual_unfairness_v2}
\end{equation}
For rank \(j\), we also define the local curvature scale as:
\begin{equation}
\small
\Lambda_j
=
\max_{e\in D(q)\setminus S_{j-1}}
\frac{\sigma}{2}
\left(
A-R(e)^2
\right)
\delta_j(e)^2 .
\label{eq:didrf_local_scale_v2}
\end{equation}
Motivated by scale matching, we compare the residual unfairness scale with the
local second-order curvature scale. When these two scales are comparable, the
policy should largely preserve the raw Taylor fairness gain; otherwise, it
should strengthen or weaken the correction according to the current income
state. Since both \(\mathcal{U}_{j-1}\) and \(\Lambda_j\) are quadratic-scale
quantities, we compare their square roots and define:
\begin{equation}
\small
\omega_j
=
\frac{1}{2}
+
\frac{
\sqrt{\mathcal{U}_{j-1}}
}{
\sqrt{\mathcal{U}_{j-1}}+\sqrt{\Lambda_j}+\epsilon
},
\label{eq:didrf_pressure_v2}
\end{equation}
where \(\epsilon\) is a small numerical constant. This factor satisfies
\(\omega_j\in[1/2,3/2)\),\footnote{The rationale and a compact proof of
boundedness, monotonicity, and sign preservation are given in
\S~\ref{app:omega-proof}.} so the calibration remains bounded.
It is close to \(1\) when the residual unfairness scale and the local curvature
scale are comparable, increases when the current income state is still highly
unfair, and decreases when the residual unfairness is already small. Therefore,
\(\omega_j\) calibrates the strength of the fairness correction without
introducing an additional trade-off hyperparameter.

The final DIDRF score for an unselected item \(d\) at rank \(j\) is:
\begin{equation}
\small
\mathrm{score}_j(d)
=
G^{\mathrm{eff}}_j(d)
+
\gamma \omega_j G^{\mathrm{fair}}_j(d)
=
p_jR^+(d)
+
\gamma \omega_j G^{\mathrm{fair}}_j(d).
\label{eq:didrf_score_v2}
\end{equation}
DIDRF then selects the item with the largest score:
\begin{equation}
\small
\pi_n[j]
=
\arg\max_{d\in D(q)\setminus S_{j-1}}
\mathrm{score}_j(d).
\label{eq:didrf_greedy_choice_v2}
\end{equation}

After selecting \(d_j=\pi_n[j]\), DIDRF updates the selected set and the prefix
interaction state:
\begin{equation}
\small
S_j=S_{j-1}\cup\{d_j\},
\qquad
B_j=B_{j-1}+R(d_j)\delta_j(d_j).
\label{eq:didrf_prefix_update_v2}
\end{equation}
The temporary income state is updated as:
\begin{equation}
\small
I^{(j)}(d,n)
=
I^{(j-1)}(d,n)
+
\delta_j(d_j)\mathbb{I}\!\left(d=d_j\right).
\label{eq:didrf_temporary_income_update_v2}
\end{equation}
The updated temporary income state is used to compute the residual unfairness
for the next rank. In implementation, the residual unfairness can be evaluated by Eq.~\eqref{eq:income_unfairness_scalar_sum}, without enumerating
all item pairs.

\begin{algorithm}[t]
\small
\caption{DIDRF Ranking Policy}
\label{alg:didrf_v2}
\begin{algorithmic}
\Require Candidate set \(D(q)\), relevance signals \(R(d)\) and \(R^+(d)\), cumulative income \(I(d,n-1)\), unit income \(\mu(d,q,\xi_n)\), position bias \(p_1,\ldots,p_{k_c}\), trade-off coefficient \(\gamma\), numerical constant \(\epsilon\).
\Ensure Ranking list \(\pi_n\).

\State Initialize \(S_0\gets\emptyset\), \(B_0\gets 0\), and \(I^{(0)}(d,n)\gets I(d,n-1)\) for all \(d\in D(q)\).
\State Compute \(\sigma\), \(A=\sum_{e\in D(q)}R(e)^2\), and \(g(d,n)\) from the pre-ranking income state \(I(d,n-1)\).

\For{\(j=1,\ldots,k_c\)}
    \State Compute \(\mathcal{U}_{j-1}=\mathrm{unfair}_{\mathrm{inc}}\!\left(q; I^{(j-1)}\right)\).
    \State Compute \(\Lambda_j\) and \(\omega_j\) by Eqs.~\eqref{eq:didrf_local_scale_v2}--\eqref{eq:didrf_pressure_v2}.
    \ForAll{\(d\in D(q)\setminus S_{j-1}\)}
        \State Compute \(G^{\mathrm{fair}}_j(d)\) by Eq.~\eqref{eq:didrf_fair_gain_v2}.
        \State Compute \(\mathrm{score}_j(d)\) by Eq.~\eqref{eq:didrf_score_v2}.
    \EndFor
    \State Select \(d_j\gets \operatorname*{arg\,max}_{d\in D(q)\setminus S_{j-1}}\mathrm{score}_j(d)\), and set \(\pi_n[j]\gets d_j\).
    \State Update \(S_j\), \(B_j\), and \(I^{(j)}(d,n)\) by Eqs.~\eqref{eq:didrf_prefix_update_v2}--\eqref{eq:didrf_temporary_income_update_v2}.
\EndFor

\State Set \(I(d,n)\gets I^{(k_c)}(d,n)\) for all \(d\in D(q)\).
\State \Return \(\pi_n\).
\end{algorithmic}
\end{algorithm}

\subsection{Offline and Online Instantiations}
\label{sec:didrf_offline_online_v2}

DIDRF uses the same ranking policy in both offline and online settings; the
only difference lies in how the relevance signals are instantiated. The
income-fairness component uses \(R(d)\), whereas the effectiveness component
uses \(R^+(d)\).

\paragraph{\textbf{Offline instantiation.}}
In the offline setting, ground-truth relevance labels are available. We set both
signals to the oracle label:
\begin{equation}
\small
R(d)=R^+(d)=R_{\mathrm{true}}(d).
\label{eq:didrf_offline_relevance_v2}
\end{equation}
This setting isolates the behavior of the ranking policy from relevance
estimation error.

\paragraph{\textbf{Online instantiation.}}
In the online setting, true relevance is not directly observed and must be
estimated from user feedback. Following online and counterfactual
learning-to-rank studies that use click and exposure feedback
~\cite{oosterhuis2021unifying}, let \(C_d^-\) and \(O_d^-\)
denote the accumulated clicks and observed exposure of item \(d\) before the
current timestep. Let \(r_q^-\) be the empirical click rate of query \(q\), with
a fixed prior used when no previous exposure has been observed. We estimate item
relevance by:
\begin{equation}
\small
\widehat R(d,n)
=
\frac{C_d^-+\lambda_R r_q^-}
{O_d^-+\lambda_R},
\label{eq:didrf_online_relevance_v2}
\end{equation}
where \(\lambda_R\) controls the shrinkage strength toward the query-level
prior. DIDRF uses this estimate as the relevance entitlement for income
fairness:
\begin{equation}
\small
R(d)\leftarrow \widehat R(d,n).
\label{eq:didrf_online_fair_signal_v2}
\end{equation}
For effectiveness, DIDRF uses an optimistic relevance signal:
\begin{equation}
\small
R^+(d)
=
\widehat R(d,n)
+
\sqrt{
\frac{\widehat R(d,n)(1-\widehat R(d,n))}
{O_d^-+1}
}.
\label{eq:didrf_online_utility_v2}
\end{equation}
This follows the optimism-under-uncertainty principle in UCB-style online
learning~\cite{auer2002finite,li2010contextual}: when relevance is uncertain,
the ranking policy should temporarily favor items whose potential effectiveness
may be underestimated. The bonus is variance-normalized, following the idea of
variance-aware confidence bonuses~\cite{audibert2009exploration}; it is larger
when the Bernoulli click estimate is uncertain and decreases as observed
exposure accumulates. We use \(R^+(d)\) only as an optimistic effectiveness
signal, not as a calibrated relevance estimate. Income fairness is computed
with \(\widehat R(d,n)\), so that exploration uncertainty is not treated as
relevance entitlement.
In both settings, DIDRF takes \(\mu(d,q,\xi_n)\) as an external unit-income
signal, which can be supplied by income simulation or estimation methods when
needed~\cite{gasior2024assessing,toder2024use,bronka2025simpaths}.

\subsection{Time Complexity Analysis}
\label{sec:didrf_complexity_v2}

For a query with \(m\) candidate items and cutoff \(k_c\), DIDRF first computes
the scalar sums needed for income-unfairness evaluation and the gradient
\(g(d,n)\) for all candidates. This preprocessing step takes \(O(m)\) time.
At each rank \(j\), DIDRF scans the remaining candidates a constant number of
times to compute \(\Lambda_j\), evaluate the score in
Eq.~\eqref{eq:didrf_score_v2}, and select the best item. After each selection,
the prefix state \(B_j\) and the scalar sums for residual income unfairness are
updated incrementally. Therefore, the total ranking-time complexity is
\(O(k_c m)\). The memory cost is \(O(m)\), mainly for storing item-level
relevance, income, unit-income, gradient, and temporary income values.
Although Eq.~\eqref{eq:income_unfairness} is pairwise, DIDRF avoids
enumerating \(O(m^2)\) item pairs by using scalar sums, \(g(d,n)\), and the
prefix state \(B_j\). This yields linear memory and \(O(k_c m)\) ranking time,
which is near-linear in the candidate-set size for a small cutoff \(k_c\).

% \subsection{Time Complexity Analysis}
% \label{sec:didrf_complexity_v2}

% For a query with \(m\) candidate items and cutoff \(k_c\), DIDRF first
% computes the scalar sums required. It also computes \(g(d,n)\) for all
% candidates. This preprocessing step takes \(O(m)\) time.

% At each rank \(j\), DIDRF scans the remaining candidates a constant number of
% times to compute \(\Lambda_j\), evaluate the score in
% Eq.~\eqref{eq:didrf_score_v2}, and select the best item. The prefix state
% \(B_j\) and the scalar sums for residual income unfairness can be updated
% incrementally after each selection. Therefore, the total ranking-time
% complexity is:
% \begin{equation}
% O(k_cm).
% \label{eq:didrf_complexity_v2}
% \end{equation}
% The memory cost is \(O(m)\), mainly for storing item-level relevance, income,
% unit-income, gradient, and temporary income values.

% Although the income-fairness metric in Eq.~\eqref{eq:income_unfairness} is
% pairwise, DIDRF never enumerates all \(O(m^2)\) item pairs during ranking.
% Using the scalar-sum form, the gradient \(g(d,n)\), and the prefix state
% \(B_j\), DIDRF summarizes the pairwise objective in linear memory and
% \(O(k_cm)\) ranking time. When \(k_c\) is treated as a small cutoff, the method
% is near-linear in the candidate-set size.

\section{Experiment}
\label{sec:experiment}

\subsection{Experimental Setup}
\label{sec:experimental_setup}

\subsubsection{Datasets}
\label{sec:experiment_dataset}

We conduct the main experiments on two standard learning-to-rank benchmarks:
MSLR-WEB30K\footnote{https://www.microsoft.com/en-us/research/project/mslr/} and MQ2008 \cite{qin2013introducing}.
MSLR-WEB30K contains 31,531 queries and 3,771,125 query-document pairs, while
MQ2008 contains 784 queries and 15,211 query-document pairs. Both datasets use
5-level relevance judgments from 0 to 4. For each query, we filter out
candidate sets with fewer than five documents and set the ranking cutoff to
\(k_c=5\). All experiments run for \(10{,}000\) ranking iterations and are
repeated with five random seeds.

\subsubsection{Baselines}
We compare DIDRF with the following methods. For fair comparison, all
fairness-aware baselines are re-derived under the proposed income-fairness
formulation: their fairness signals are computed from ranking-induced income,
while their original optimization mechanisms are preserved.
\begin{itemize}[leftmargin=1.2em, labelsep=0.4em, itemsep=2pt, topsep=2pt]
\item \textbf{RandomK}: Randomly permutes the items.
\item \textbf{TopK}: Ranks items by descending effectiveness.
\item \textbf{FairK}: Ranks items solely by income-fairness gain
\(\Delta \mathrm{fair}_{\mathrm{inc}}(q,n;\pi_n)\).
\item \textbf{FairCo} \cite{morik2020controlling}: Feedback-control fair
ranking based on accumulated fairness deviation.
\item \textbf{MMF} \cite{yang2021maximizing}: Marginal fair-ranking method that
emphasizes fairness at top positions.
\item \textbf{MCFair} \cite{yang2023marginal}: Marginal-contribution fair
ranking method using objective-gradient-based scores.
\item \textbf{FARA} \cite{yang2023fara}: Future-aware fair ranking method that
plans multiple upcoming rank lists with a Taylor-based objective.
\item \textbf{TaxRank-Income} \cite{xu2024taxation}: Income-fair adaptation of
TaxRank, treating cumulative income as the taxed resource.
\item \textbf{DIDRF}: Our derivative-aware income-fair ranking method with
prefix-aware correction and state calibration.
\end{itemize}
We tune \(\gamma\in[0.0,1000.0]\) for all fairness-aware methods except MMF,
which uses \(\gamma\in[0.0,1.0]\). MCFair and FARA additionally tune
\(\eta\in[0,100]\), which controls the exploration bonus under online relevance
uncertainty. For FARA, the future horizon is set to \(100\).

\subsubsection{Income Simulation Environments}
\label{sec:experiment_income_simulation}

Following common semi-synthetic LTR/CLTR evaluation protocols
\cite{gupta2023safe,hager2024unbiased}, we simulate the unit-income signal
\(\mu(d,q,\xi_i)\) semi-synthetically: relevance labels and candidate sets come
from standard LTR benchmarks, while the exposure-to-income process is calibrated
from real commercial logs. We construct two log-calibrated environments using
the Criteo Attribution Modeling for Bidding Dataset \cite{diemert2017attribution}
and the RecSys Challenge 2015 YOOCHOOSE dataset \cite{benshimon2015recsys},
referred to as \textbf{Criteo-Ads} and \textbf{YOOCHOOSE-Ecom}, respectively.

In Criteo-Ads, each trajectory corresponds to an advertising campaign and is
calibrated from attributed conversion value aggregated over campaign-time pairs;
after filtering and time binning, it contains 675 campaign trajectories over
62 12-hour bins. In YOOCHOOSE-Ecom, each trajectory corresponds to a product and
is calibrated from purchase revenue normalized by item-level click signals; after
filtering and time binning, it contains 50,000 item trajectories over 182
24-hour bins. These environments represent two common commercial
income-generation processes: advertising conversions and e-commerce purchases.

For both environments, we build a trajectory bank \(\mathcal{M}\) from raw
income logs using the same preprocessing pipeline, including smoothing,
logarithmic transformation, outlier clipping, and normalization to \([0,1]\). Since standard LTR benchmarks do not provide native income logs, we do
not assume semantic correspondence between LTR documents and real campaigns or
products. Instead, each query-document pair is deterministically assigned to one
log-derived trajectory:
\begin{equation}
\small
\label{eq:income_trajectory_mapping}
\phi(q,d)
=
\mathrm{Hash}(q,d,\mathrm{seed}) \bmod |\mathcal{M}|,
\qquad
\mu(d,q,\xi_i)=\mathcal{M}_{\phi(q,d)}(b_i).
\end{equation}
Here, \(b_i\) is the global time bin used at ranking iteration \(i\). We replay
each calibrated trajectory cyclically: for zero-indexed iteration \(t\),
\(b_t=\lfloor t\rfloor \bmod B\), where \(B=62\) for Criteo-Ads and \(B=182\)
for YOOCHOOSE-Ecom. Thus, 10,000 ranking iterations correspond to repeated
passes over calibrated temporal income states, not to 10,000 physical 12-hour
or 24-hour periods. Specifically, Criteo-Ads covers 161 full cycles plus 18
bins, and YOOCHOOSE-Ecom covers 54 full cycles plus 172 bins. We use no
interpolation, truncation, or random starting offset. This construction retains
the empirical temporal dynamics, sparsity, and income heterogeneity of real
advertising and e-commerce logs, while decoupling them from the semantic
identities of LTR documents.

\subsubsection{Experimental Settings}

We use two experimental settings: \textbf{offline} and \textbf{online}. In the offline setting, item relevance is predetermined or well-estimated, allowing all ranking algorithms to rely on accurate relevance values. The relevance signal for each document-query pair $(d, q)$ is converted from its relevance judgment $y$ \cite{ai2018unbiased}: 
\begin{equation}
\small
% \small
  R(d,q) = \epsilon_R + (1-\epsilon_R)\frac{2^y-1}{2^{y_{\max}}-1}
  \label{eq:offline_relevance_conversion}
\end{equation}
where $y_{\max}$ is the maximum relevance annotation and $\epsilon_R=0.1$. In the online setting, ranking optimization and relevance learning happen simultaneously, reflecting real-world conditions. Here, all algorithms use relevance estimation from
Eq.~\eqref{eq:didrf_online_relevance_v2} with \(\lambda_R=1.0\). Meanwhile, for simplicity, following prior studies \cite{morik2020controlling,oosterhuis2021unifying}, we assume the users' examination probability $p(e=1\mid d,\pi_i)$ is known. It is simulated as:
\begin{equation}
\small
\label{eq:experiment_position_bias}
% \small
p(e=1\mid d,\pi_i)=\left\{
\begin{aligned}
& \frac{1}{\log_2(\operatorname{rank}(d\mid \pi_i)+1)},\ \ if\ \operatorname{rank}(d\mid \pi_i)\le k_c\\
& 0 ,\ \ otherwise
\end{aligned}
\right.
\end{equation}
\subsubsection{Evaluation}
We evaluate algorithms on effectiveness and fairness. For effectiveness, we use the cNDCG with $\alpha=1$ and evaluation cutoffs $k\in\{1,2,3,4,5\}$:
\begin{equation}
\small
\mathrm{cNDCG}_{\mathrm{avg}}@k=\sum_{\tau=1}^{n}\alpha^{n-\tau}\mathrm{NDCG}@k(\pi_{\tau},q)
\label{eq:experiment_avg_cndcg}
\end{equation}
For fairness, we use the aggregated income unfairness from Eq.~\eqref{eq:overall_income_unfairness}. Each experiment is run five times, and the average performance on the test set is reported.

\begin{table*}[t]
\centering
\caption{Fairness capacity in offline and online settings. 
Results are reported under each method's best-fairness setting. 
Values are mean$_{(\mathrm{std})}$ over five seeds; MSLR-WEB30K unfairness is scaled by \(10^3\). 
Fairness-aware baselines are adapted to Income Fairness. 
Bold marks the best \(\mathrm{cNDCG}@5\) among fairness-aware methods and the best unfairness/time in each column; \(\dagger\) marks \(p<0.05\) under a two-sided paired \(t\)-test against the strongest non-DIDRF fairness-aware baseline.}
\label{tab:rq1_fairness_capacity_didrf_c1}
% DIDRF selected alphas: off-MQ=200, off-MSLR=1000, on-MQ=50, on-MSLR=500.

{\small
\setlength{\tabcolsep}{1.8pt}
\renewcommand{\arraystretch}{0.98}
\setlength{\aboverulesep}{0.35ex}
\setlength{\belowrulesep}{0.35ex}
\setlength{\cmidrulesep}{0.20ex}
\def\tabstd#1{_{\scriptscriptstyle(#1)}}
\def\best#1#2{\textbf{#1}\({}_{\scriptscriptstyle(#2)}\)}
\def\sig#1#2#3{#1\({}^{#3}_{\scriptscriptstyle(#2)}\)}
\def\bestsig#1#2#3{\textbf{#1}\({}^{#3}_{\scriptscriptstyle(#2)}\)}
\def\vsep{\hspace{2.5pt}\vrule width 0.25pt\hspace{2.5pt}}

\begin{tabularx}{0.96\textwidth}{
@{}c!{\vsep}c!{\vsep}l!{\vsep}
>{\centering\arraybackslash}X
>{\centering\arraybackslash}X
>{\centering\arraybackslash}X
!{\vsep}
>{\centering\arraybackslash}X
>{\centering\arraybackslash}X
>{\centering\arraybackslash}X
@{}}
\toprule[0.9pt]
\multirow{2}{*}{Setting}
& \multirow{2}{*}{\shortstack{Algorithm\\Type}}
& \multirow{2}{*}{Method}
& \multicolumn{3}{c!{\vsep}}{MQ2008 / Criteo-Ads}
& \multicolumn{3}{c@{}}{MSLR-WEB30K / YOOCHOOSE-Ecom} \\
\cmidrule(lr){4-6}\cmidrule(lr){7-9}
& &
& \(\mathrm{cNDCG}@5\) & Unfairness & Time(s)
& \(\mathrm{cNDCG}@5\) & Unfair. \(\times 10^3\) & Time(s) \\
\midrule

\multirow{9}{*}{Offline}
& \multirow{2}{*}{\shortstack{Fairness-\\agnostic}}
& TopK
& $200.00\tabstd{0.00}$
& $24.18\tabstd{0.38}$
& $2.40\tabstd{0.03}$
& $200.00\tabstd{0.00}$
& $3.04\tabstd{0.04}$
& $7.04\tabstd{0.09}$ \\
& & RandomK
& $106.98\tabstd{2.74}$
& $25.10\tabstd{0.81}$
& \best{2.36}{0.01}
& $79.01\tabstd{1.89}$
& $1.73\tabstd{0.05}$
& \best{6.92}{0.05} \\

\cmidrule(lr){2-9}

& \multirow{7}{*}{\shortstack{Fairness-\\aware}}
& FairCo
& $172.24\tabstd{2.01}$
& $2.22\tabstd{0.09}$
& $3.14\tabstd{0.01}$
& $101.78\tabstd{2.24}$
& $1.14\tabstd{0.01}$
& $8.13\tabstd{0.24}$ \\
& & FairK
& $171.73\tabstd{2.68}$
& $1.92\tabstd{0.08}$
& $2.68\tabstd{0.02}$
& $138.32\tabstd{2.09}$
& $1.07\tabstd{0.02}$
& $7.66\tabstd{0.09}$ \\
& & MMF
& $160.62\tabstd{3.18}$
& $7.09\tabstd{0.18}$
& $28.70\tabstd{0.33}$
& $78.87\tabstd{1.06}$
& $1.59\tabstd{0.03}$
& $136.50\tabstd{0.67}$ \\
& & MCFair
& $171.75\tabstd{2.14}$
& $1.34\tabstd{0.05}$
& $3.08\tabstd{0.01}$
& $130.27\tabstd{2.67}$
& $1.10\tabstd{0.01}$
& $7.92\tabstd{0.10}$ \\
& & FARA
& $173.22\tabstd{2.81}$
& $2.57\tabstd{0.08}$
& $15.41\tabstd{1.01}$
& $150.78\tabstd{2.38}$
& $1.46\tabstd{0.01}$
& $12.50\tabstd{0.06}$ \\
& & TaxRank-Income
& $175.52\tabstd{1.52}$
& $6.11\tabstd{0.14}$
& $4.17\tabstd{0.06}$
& $113.44\tabstd{1.65}$
& $1.24\tabstd{0.01}$
& $8.80\tabstd{0.04}$ \\
& & DIDRF (Ours)
& \bestsig{176.57}{2.10}{\dagger}
& \bestsig{1.21}{0.06}{\dagger}
& $4.83\tabstd{0.02}$
& \bestsig{177.74}{1.01}{\dagger}
& \bestsig{1.01}{0.01}{\dagger}
& $8.88\tabstd{0.04}$ \\

\midrule

\multirow{9}{*}{Online}
& \multirow{2}{*}{\shortstack{Fairness-\\agnostic}}
& TopK
& $197.25\tabstd{0.89}$
& $13.59\tabstd{0.17}$
& $2.38\tabstd{0.01}$
& $102.32\tabstd{1.23}$
& $2.86\tabstd{0.07}$
& \best{6.83}{0.05} \\
& & RandomK
& $104.48\tabstd{1.15}$
& $25.59\tabstd{0.89}$
& \best{2.37}{0.02}
& $78.95\tabstd{1.54}$
& $1.73\tabstd{0.06}$
& $7.13\tabstd{0.21}$ \\

\cmidrule(lr){2-9}

& \multirow{7}{*}{\shortstack{Fairness-\\aware}}
& FairCo
& $170.36\tabstd{2.96}$
& $2.45\tabstd{0.11}$
& $3.12\tabstd{0.02}$
& $79.25\tabstd{1.18}$
& $1.49\tabstd{0.03}$
& $8.37\tabstd{0.16}$ \\
& & FairK
& $170.13\tabstd{3.00}$
& $2.22\tabstd{0.07}$
& $2.67\tabstd{0.01}$
& $83.60\tabstd{1.23}$
& $1.46\tabstd{0.02}$
& $7.71\tabstd{0.17}$ \\
& & MMF
& $158.42\tabstd{0.88}$
& $7.18\tabstd{0.09}$
& $28.56\tabstd{0.22}$
& $79.59\tabstd{1.14}$
& $1.59\tabstd{0.02}$
& $136.34\tabstd{0.35}$ \\
& & MCFair
& $169.47\tabstd{1.98}$
& $1.69\tabstd{0.05}$
& $3.05\tabstd{0.02}$
& $83.20\tabstd{1.24}$
& $1.65\tabstd{0.03}$
& $7.93\tabstd{0.16}$ \\
& & FARA
& $171.05\tabstd{2.70}$
& $3.79\tabstd{0.17}$
& $16.16\tabstd{0.24}$
& $79.04\tabstd{1.27}$
& $1.60\tabstd{0.04}$
& $12.03\tabstd{0.07}$ \\
& & TaxRank-Income
& $172.58\tabstd{1.84}$
& $6.55\tabstd{0.13}$
& $4.15\tabstd{0.03}$
& $78.54\tabstd{1.69}$
& $1.61\tabstd{0.02}$
& $8.93\tabstd{0.07}$ \\
& & DIDRF (Ours)
& \bestsig{176.73}{2.53}{\dagger}
& \bestsig{1.50}{0.05}{\dagger}
& $4.21\tabstd{1.08}$
& \bestsig{89.83}{1.13}{\dagger}
& \bestsig{1.35}{0.02}{\dagger}
& $8.79\tabstd{0.10}$ \\

\bottomrule[0.9pt]
\end{tabularx}
}
\end{table*}

\subsection{Results and Analysis}
\label{sec:experiment_results}

\subsubsection{What level of income fairness does DIDRF empirically achieve?}
\label{sec:experiment_main_findings}

Table~\ref{tab:rq1_fairness_capacity_didrf_c1}\footnote{The complete experiment covers two datasets, two income environments, and two relevance settings. Due to space constraints, we report representative settings that cover different data scales, income scenarios, and offline/online conditions. Across the complete set of settings, DIDRF achieves significantly best income fairness and \(\mathrm{cNDCG@5}\) among fairness-aware methods.}
shows that DIDRF achieves the
lowest income unfairness across scenarios and settings, with
statistically significant improvements over the strongest baselines.
This result is consistent with DIDRF's marginal income-correction mechanism:
instead of fixed exposure quotas or simple low-income boosting, DIDRF adjusts
each candidate by its effectiveness signal, current income state, and prefix
interaction. The correction is stronger for under-compensated items and weaker
when residual income unfairness is small, enabling targeted income-fairness
correction during ranking.

The comparison with the baselines further supports this interpretation.
MCFair and FairK also use marginal or fairness-aware ranking signals, but their
corrections are more limited: MCFair operates at a coarser granularity, while
FairK focuses more on fairness signals and considers the dynamic tradeoff with
relevance less explicitly. FARA introduces future allocation planning, but its
planned target may not fully reflect the current income state. TaxRank-Income
mainly applies welfare-style low-income compensation, and MMF emphasizes
protecting the least-served items. 
% In contrast, DIDRF uses a series of mechanisms to adjust ranking decisions in a more comprehensive and
% fine-grained manner, leading to stronger income-fairness performance across
% different settings.

\begin{figure}[t]
\centering

\subfloat[MQ2008 / YOOCHOOSE-Ecom / Offline]{
    \includegraphics[width=0.95\linewidth]{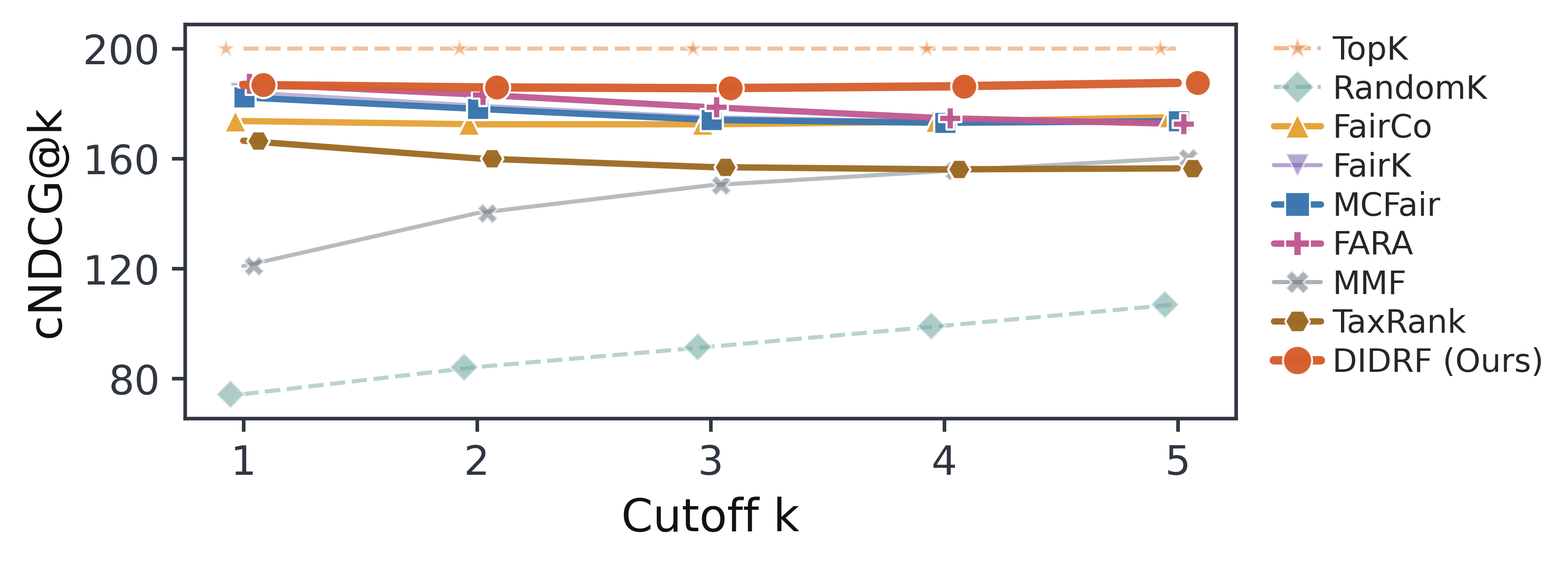}
}
\vspace{-1.0em}
\subfloat[MSLR-WEB30K / Criteo-Ads / Online]{
    \includegraphics[width=0.95\linewidth]{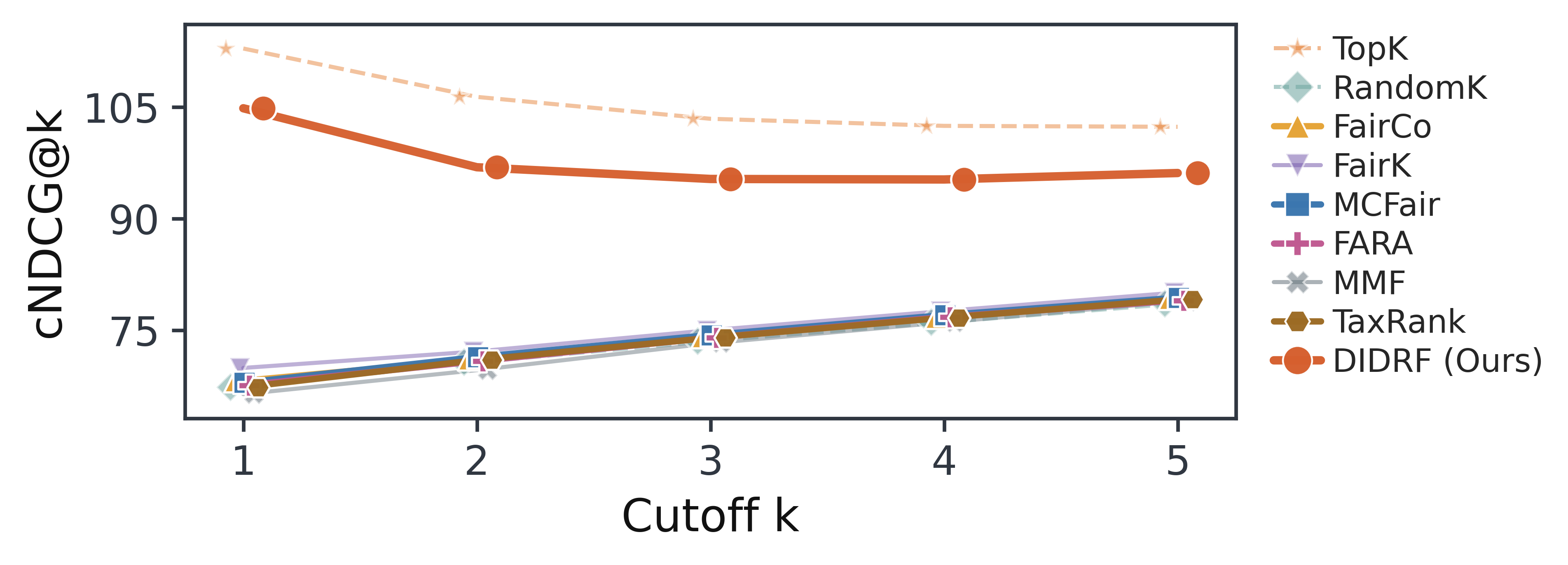}
}

\caption{\(\mathrm{cNDCG}@k_c\) across ranking cutoffs from 1 to 5.}
\label{fig:rq2}
\vspace{-1.5em}
\end{figure}

\begin{figure*}[t]
\centering
\captionsetup[subfloat]{justification=centering,singlelinecheck=false}

\subfloat[MQ2008 / YOOCHOOSE-Ecom /\protect\\Offline]{
    \includegraphics[width=0.24\linewidth]{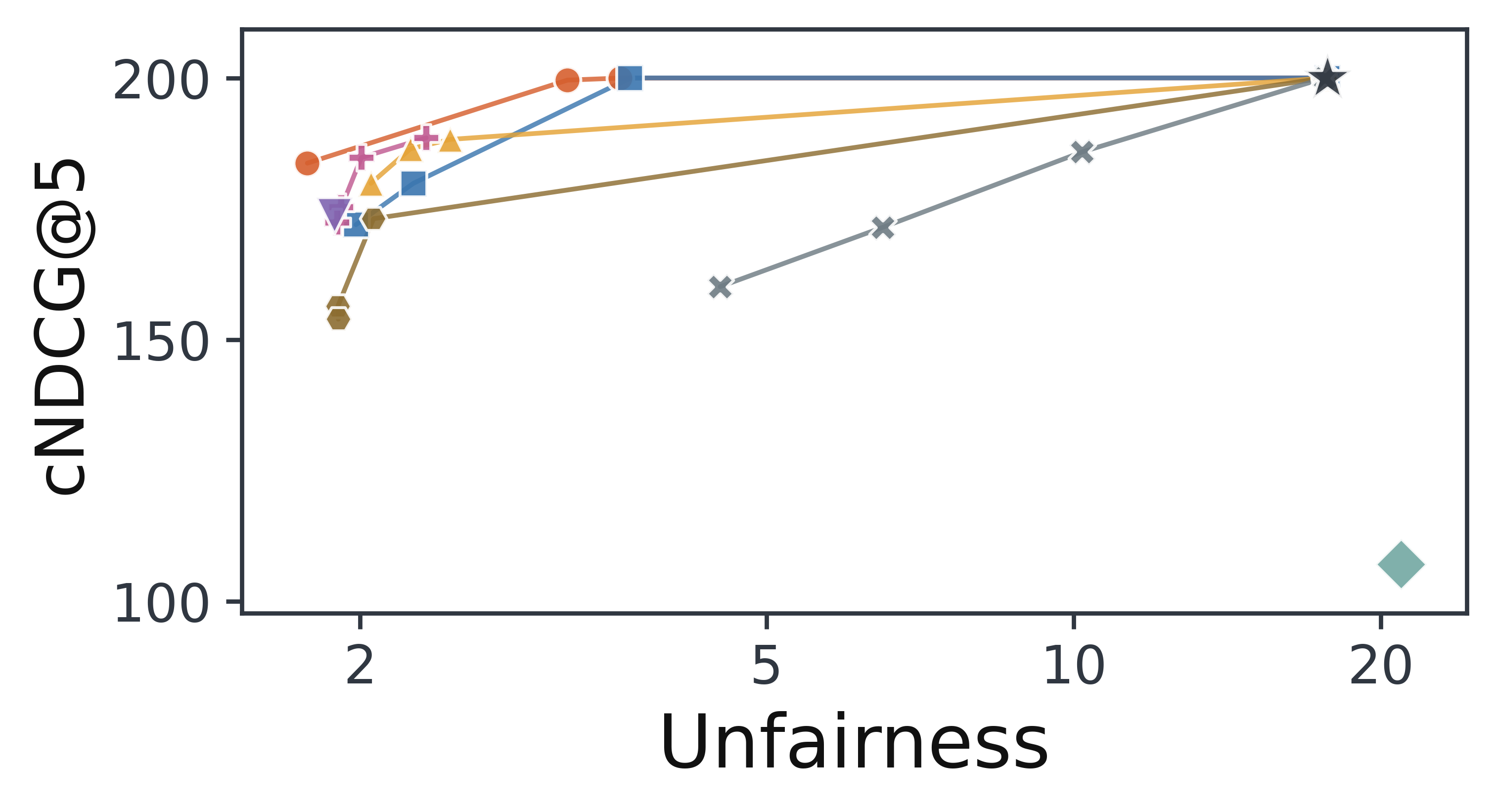}
}
\hfill
\subfloat[MQ2008 / YOOCHOOSE-Ecom /\protect\\Online]{
    \includegraphics[width=0.24\linewidth]{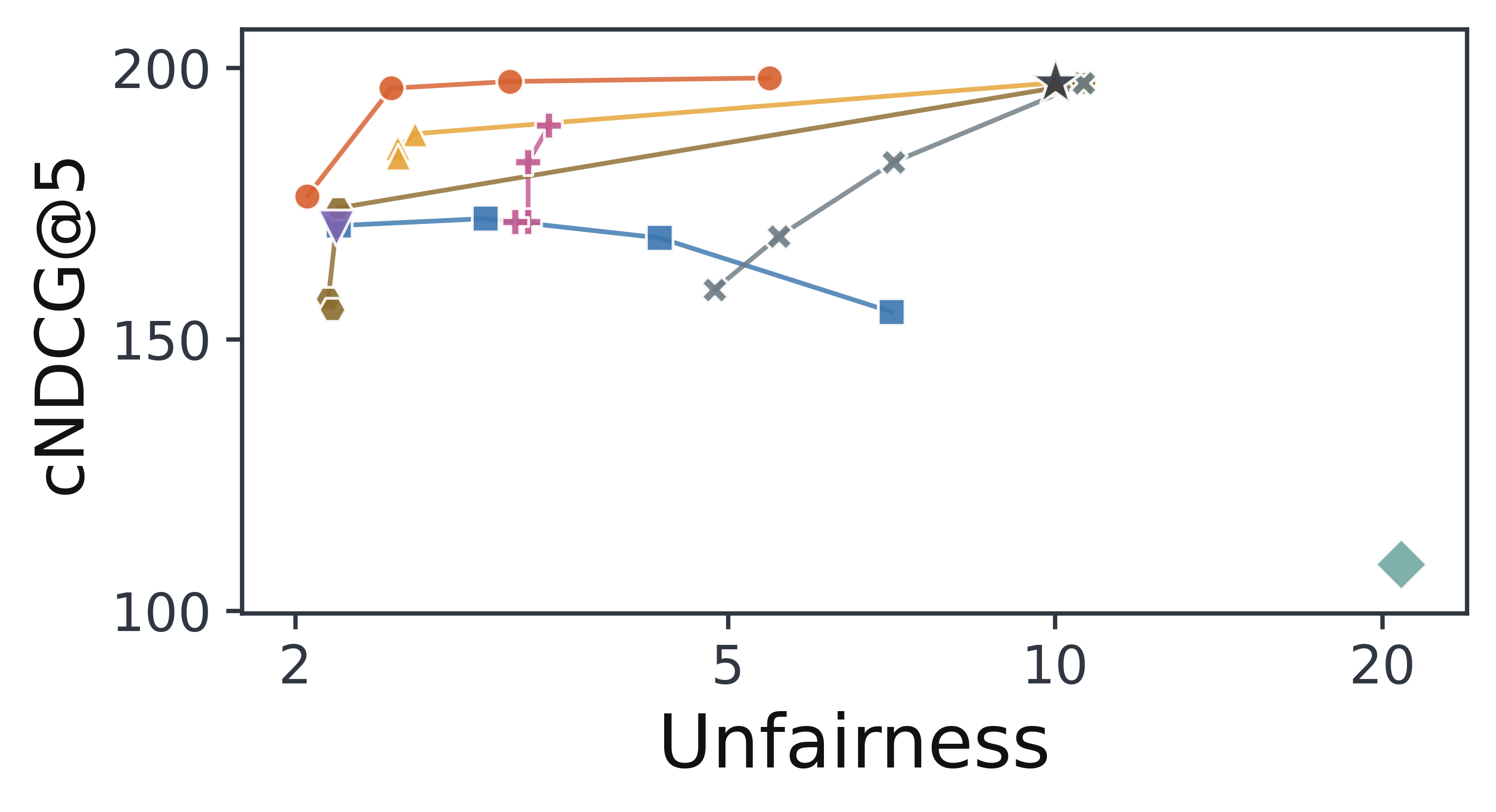}
}
\hfill
\subfloat[MSLR-WEB30K / Criteo-Ads /\protect\\Offline]{
    \includegraphics[width=0.24\linewidth]{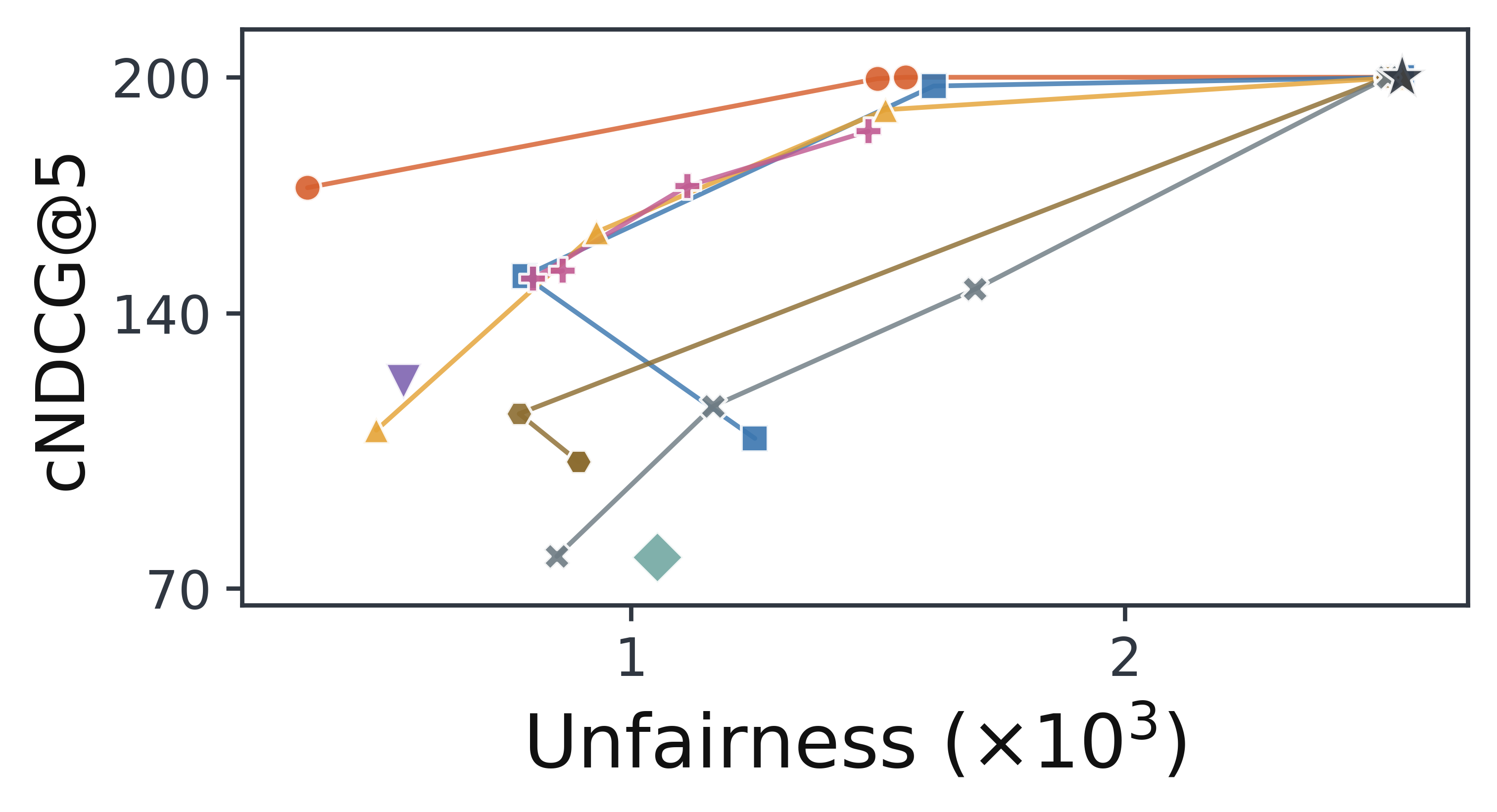}
}
\hfill
\subfloat[MSLR-WEB30K / Criteo-Ads /\protect\\Online]{
    \includegraphics[width=0.24\linewidth]{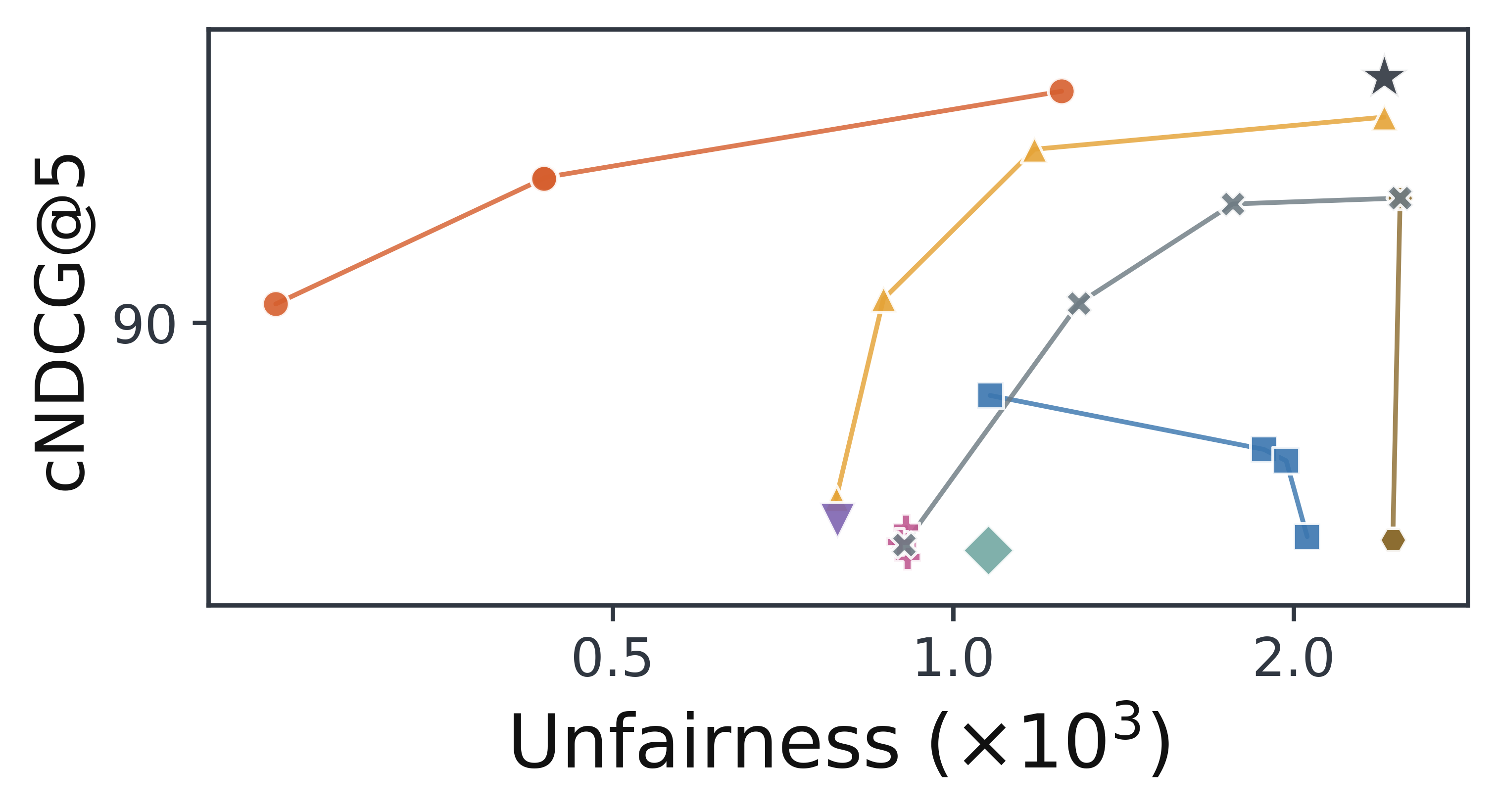}
}

\vspace{0.35em}

\includegraphics[width=0.58\linewidth]{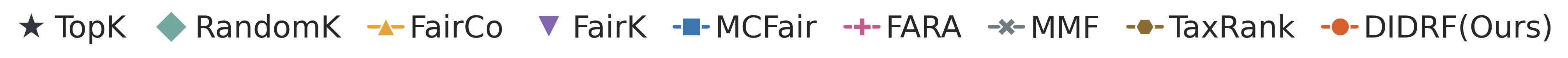}

\caption{Fairness--effectiveness tradeoff curves across four representative
settings. Lower income unfairness and higher \(\mathrm{cNDCG}@5\) indicate
better tradeoffs.}
\label{fig_8}
\vspace{-1.5em}
\end{figure*}

\subsubsection{How effective is DIDRF at different cutoffs?}
\label{sec:experiment_tradeoff_analysis}

Figure~\ref{fig:rq2} shows that DIDRF maintains competitive effectiveness among fairness-aware methods across cutoffs while achieving strong income-fairness performance.
This behavior is consistent with DIDRF's design: fairness is used as a
state-dependent correction rather than a replacement for relevance. During
ranking, the effectiveness signal remains part of every rank decision, while
the fairness correction is adjusted according to the current income state, the
selected prefix, and the remaining income unfairness. In online settings, the
optimistic relevance signal further helps under-observed but potentially
relevant items remain competitive. These mechanisms allow DIDRF to improve
income fairness without overly sacrificing ranking effectiveness, leading to
stable \(\mathrm{cNDCG}@k\) across cutoffs.
The trend is most pronounced in the MSLR-WEB30K / Criteo-Ads / Online setting,
suggesting that DIDRF remains robust under larger candidate sets, estimated
relevance, and heterogeneous unit income.

\subsubsection{Can DIDRF achieve a better effectiveness--fairness balance?}
\label{sec:experiment_pareto_tradeoff}

Figure~\ref{fig_8} shows the empirical fairness--effectiveness frontiers in
four representative settings, covering two data pairs under both offline and
online conditions. Lower income unfairness and higher \(\mathrm{cNDCG}@5\)
indicate better tradeoffs. Across the four panels, DIDRF yields a stronger
frontier than the baselines: it reaches lower income unfairness while preserving
competitive \(\mathrm{cNDCG}@5\), rather than improving fairness mainly through
a large effectiveness loss.
This tradeoff is consistent with DIDRF's dynamic correction design. Relevance
remains a core ranking signal, while income fairness is introduced as a
state-dependent correction based on the current income distribution. The
marginal income correction, prefix interaction, and state calibration together
allow DIDRF to adjust ranking decisions according to the current fairness state.
As a result, DIDRF better balances relevance preservation and income-fairness
improvement across operating points.

\subsubsection{Is DIDRF efficient enough for practical deployment?}
\label{sec:experiment_efficiency}

Table~\ref{tab:rq1_fairness_capacity_didrf_c1} reports the
simulator-recorded runtime under offline and online settings. FairK and MCFair
are the fastest fairness-aware baselines in our experiments, and DIDRF remains
close to them while being much faster than methods with heavier planning or
max-min allocation. This result is consistent with the complexity analysis in
\S\ref{sec:didrf_complexity_v2}. Under the fixed top-\(k\) protocol, DIDRF has
the same \(O(m)\) asymptotic order as FairK and MCFair, where \(m\) is the
candidate-set size. Its slightly higher runtime mainly comes from the constant
overhead of income-aware state updates, including residual-unfairness and
prefix-state updates. Thus, DIDRF remains computationally practical for
deployment-oriented ranking settings while providing stronger income-fairness
performance.

\begin{table}[t]
\centering
\caption{Results under the exposure fairness setting. Cells report
mean$_{\mathrm{(std)}}$ over five seeds at each method's best-fairness setting;
MSLR-WEB30K unfairness is scaled by \(10^3\). Bold marks the best
\(\mathrm{cNDCG}@5\) among fairness-aware methods and the best unfairness; \(\dagger\) marks \(p<0.05\) under a two-sided paired \(t\)-test against the strongest non-DIDRF fairness-aware baseline.}
\label{tab:rq4_fd1_exposure_control}
% DIDRF selected alphas: MQ2008/offline = 5, MSLR-WEB30k/online = 500.

{\small
\setlength{\tabcolsep}{2.1pt}
\renewcommand{\arraystretch}{0.98}
\setlength{\aboverulesep}{0.35ex}
\setlength{\belowrulesep}{0.35ex}
\setlength{\cmidrulesep}{0.20ex}
\def\tabstd#1{_{\scriptscriptstyle(#1)}}
\def\sig#1#2#3{#1\({}^{#3}_{\scriptscriptstyle(#2)}\)}
\def\bestsig#1#2#3{\textbf{#1}\({}^{#3}_{\scriptscriptstyle(#2)}\)}
\def\vsep{\hspace{2.0pt}\vrule width 0.25pt\hspace{2.0pt}}

\begin{tabularx}{\columnwidth}{
@{}l!{\vsep}
>{\centering\arraybackslash}X
>{\centering\arraybackslash}X
!{\vsep}
>{\centering\arraybackslash}X
>{\centering\arraybackslash}X
@{}}
\toprule[0.9pt]
\multirow{2}{*}{Method}
& \multicolumn{2}{c!{\vsep}}{\shortstack{MQ2008\\Offline}}
& \multicolumn{2}{c@{}}{\shortstack{MSLR-WEB30K\\Online}} \\
\cmidrule(lr){2-3}\cmidrule(lr){4-5}
& \(\mathrm{cNDCG}@5\) & Unfairness
& \(\mathrm{cNDCG}@5\) & Unfair. \(\times 10^3\) \\
\midrule

TopK
& $200.00\tabstd{0.00}$
& $115.26\tabstd{3.47}$
& $102.32\tabstd{1.23}$
& $13.54\tabstd{0.23}$ \\
RandomK
& $106.68\tabstd{2.08}$
& $135.94\tabstd{2.92}$
& $75.51\tabstd{0.57}$
& $5.83\tabstd{0.14}$ \\

\cmidrule(lr){1-5}

FairCo
& $166.85\tabstd{1.99}$
& $12.22\tabstd{0.51}$
& $76.47\tabstd{1.15}$
& $4.49\tabstd{0.03}$ \\
FairK
& $167.96\tabstd{2.46}$
& $11.97\tabstd{0.51}$
& $75.78\tabstd{0.83}$
& $4.48\tabstd{0.03}$ \\
MMF
& $160.10\tabstd{3.31}$
& $27.54\tabstd{0.78}$
& $76.84\tabstd{0.72}$
& $4.86\tabstd{0.05}$ \\
MCFair
& $168.94\tabstd{2.57}$
& $11.86\tabstd{0.51}$
& $75.64\tabstd{1.07}$
& $4.47\tabstd{0.03}$ \\
FARA
& $170.95\tabstd{2.59}$
& $11.97\tabstd{0.49}$
& $76.43\tabstd{1.52}$
& $4.89\tabstd{0.09}$ \\
TaxRank
& $168.93\tabstd{2.07}$
& $12.19\tabstd{0.51}$
& $76.15\tabstd{1.68}$
& $4.53\tabstd{0.05}$ \\
DIDRF (Ours)
& \bestsig{172.43}{3.41}{\dagger}
& \bestsig{11.55}{0.51}{\dagger}
& \bestsig{79.51}{1.47}{\dagger}
& \bestsig{4.31}{0.04}{\dagger} \\
\bottomrule[0.9pt]
\end{tabularx}
}
\vspace{-1.5em}
\end{table}

\subsubsection{How does DIDRF perform when income fairness degenerates to exposure fairness?}
\label{sec:experiment_exposure_fairness}

Table~\ref{tab:rq4_fd1_exposure_control} reports the results when income fairness
degenerates to exposure fairness. With homogeneous unit income, ranking-induced
income becomes proportional to exposure. DIDRF still achieves the best observed
exposure-fairness performance among the evaluated methods, showing that its
effectiveness is not limited to heterogeneous-income settings. This result suggests that DIDRF naturally covers the classical exposure-fairness
case. When income differs from exposure only by a constant scale, its
income-fairness correction reduces to exposure-fairness correction. Therefore,
DIDRF can continue to reduce disparity while preserving relevance
in the ranking decision.

\begin{figure}[t]
\centering
\captionsetup[subfloat]{justification=centering,singlelinecheck=false}

\subfloat[MQ2008 / YOOCHOOSE-Ecom /\protect\\Online]{
    \includegraphics[width=0.48\linewidth]{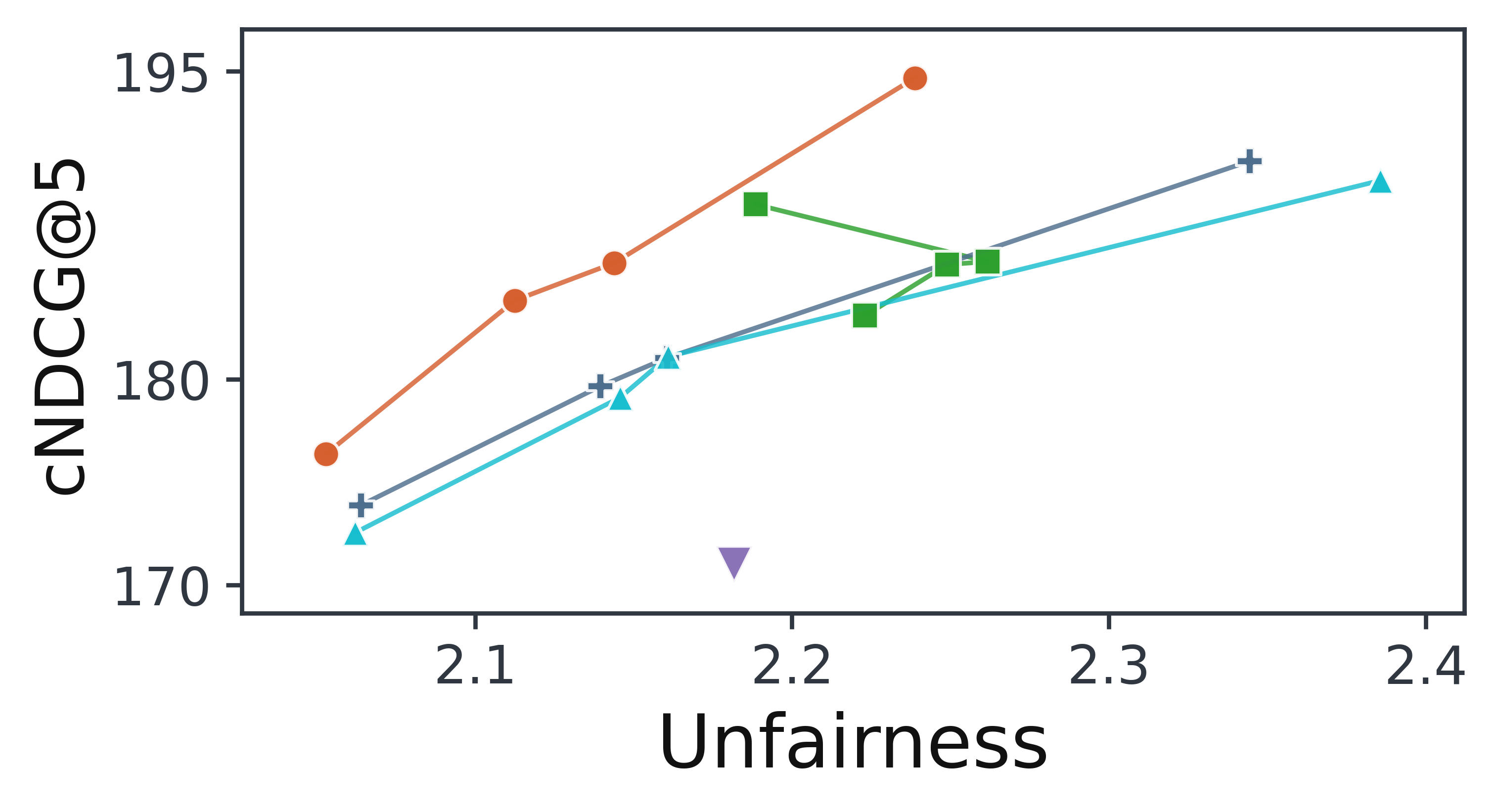}
}
\subfloat[MSLR-WEB30K / Criteo-Ads /\protect\\Offline]{
    \includegraphics[width=0.48\linewidth]{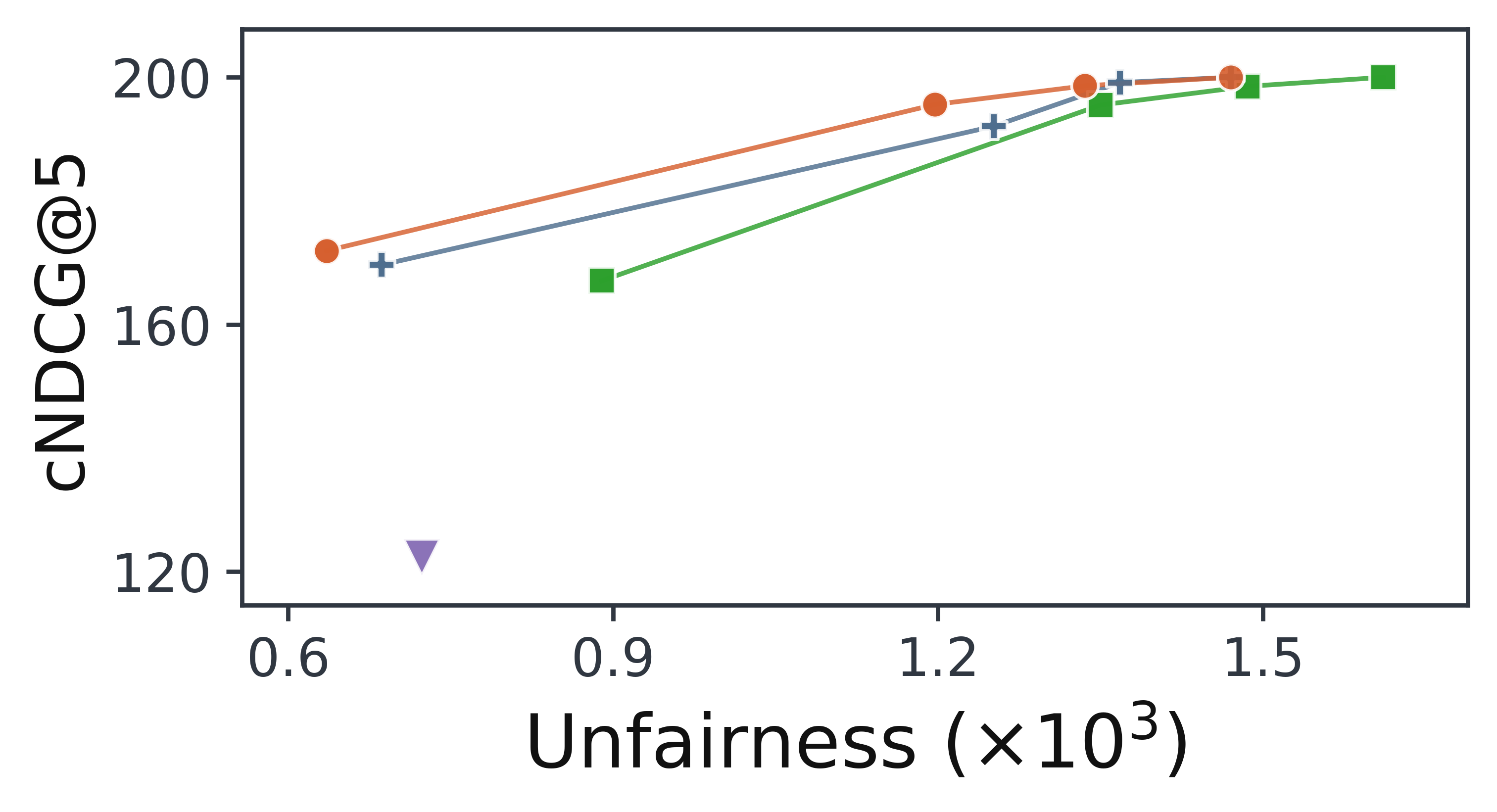}
}

\vspace{0.35em}

\includegraphics[width=0.9\linewidth]{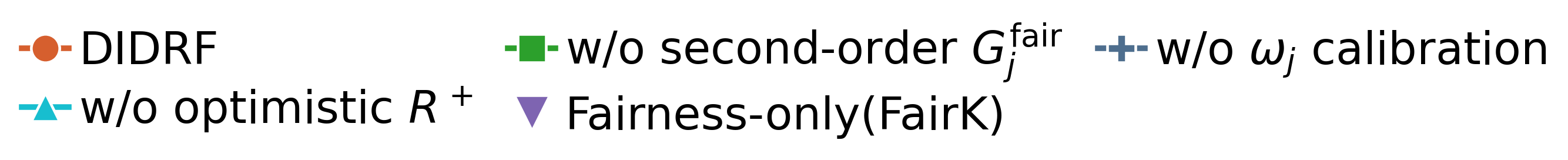}

\caption{Fairness--effectiveness Pareto curves for ablation.}
\label{fig:rq6_ablation}
\vspace{-2.0em}
\end{figure}

\subsubsection{Which components are responsible for DIDRF's performance?}
\label{sec:experiment_ablation}

Figure~\ref{fig:rq6_ablation} compares the full DIDRF with variants that remove
the second-order fairness term, state calibration, or optimistic relevance
signal, as well as a fairness-only reference. The full model achieves the most
favorable fairness--effectiveness balance in both representative settings:
it preserves higher \(\mathrm{cNDCG}@5\) at comparable or lower income
unfairness, whereas each ablated variant leads to a weaker or less stable
tradeoff. These results support the contribution of DIDRF's main components. The
second-order, prefix-aware correction captures the listwise effect of each
placement; state calibration adjusts the correction strength to the current
income state; and the optimistic \(R^+\) signal improves robustness in the
online setting. The fairness-only reference further confirms that removing
relevance from the ranking decision causes substantial effectiveness loss.
Overall, DIDRF benefits from combining relevance-aware ranking with dynamic
income-fairness correction.

\section{CONCLUSION}
This paper studies fair ranking when provider income is not determined by
exposure alone. We formalize income fairness and propose DIDRF, a dynamic
income-derivative-aware ranking algorithm that applies marginal, state-aware
corrections in both offline and online settings. Experiments in
advertising- and e-commerce-calibrated environments show that DIDRF consistently
improves income fairness while maintaining competitive ranking effectiveness.
These results highlight the importance of optimizing realized provider utility
rather than relying only on exposure as a proxy. Future work will study
deployment with learned income predictors and real production logs.

\appendix

\section{Derivation of Candidate-Dependent Fairness Gain}
\label{app:fair-gain-derivation}

We derive the candidate-dependent fairness gain used in
Eq.~\eqref{eq:didrf_fair_gain_v2}. At rank \(j\), suppose the prefix
\(\pi_n[1],\ldots,\pi_n[j-1]\) is fixed and candidate \(d\notin S_{j-1}\) is
considered. Let \(\delta_d=\delta_j(d)\) and
\(\delta_s=\delta_s(\pi_n[s])\) for \(s<j\). Recall that
\(A=\sum_{e\in D(q)}R(e)^2\) and
\(B_{j-1}=\sum_{s<j}R(\pi_n[s])\delta_s(\pi_n[s])\).

From Eq.~\eqref{eq:didrf_fairness_taylor_v2}, the candidate-dependent part is:
\begin{equation}
\small
\begin{aligned}
G^{\mathrm{fair}}_j(d)
&=
g(d,n)\delta_d
+
\sum_{s<j}H_{d,\pi_n[s]}\delta_d\delta_s
+
\frac{1}{2}H_{d,d}\delta_d^2,
\\
H_{x,y}
&=
-\sigma
\left(
\mathbb{I}(x=y)A-R(x)R(y)
\right).
\end{aligned}
\end{equation}
Since \(d\notin S_{j-1}\), we have
\(H_{d,\pi_n[s]}=\sigma R(d)R(\pi_n[s])\) for every prefix item and
\(H_{d,d}=-\sigma(A-R(d)^2)\). Therefore,
\begin{equation}
\small
\begin{aligned}
\sum_{s<j}H_{d,\pi_n[s]}\delta_d\delta_s
&=
\sigma R(d)\delta_d
\sum_{s<j}R(\pi_n[s])\delta_s(\pi_n[s])
=
\sigma R(d)B_{j-1}\delta_d,
\\
\frac{1}{2}H_{d,d}\delta_d^2
&=
-\frac{\sigma}{2}
\left(A-R(d)^2\right)\delta_d^2.
\end{aligned}
\end{equation}
Combining the first-order term, prefix interaction term, and self-curvature
term gives:
\begin{equation}
\small
G^{\mathrm{fair}}_j(d)
=
\delta_j(d)
\left[
g(d,n)+\sigma R(d)B_{j-1}
\right]
-
\frac{\sigma}{2}
\left(
A-R(d)^2
\right)
\delta_j(d)^2.
\end{equation}
Terms depending only on the fixed prefix are omitted because they are constant
with respect to the current candidate and do not affect the greedy choice at
rank \(j\).

\section{Rationale and Properties of State Calibration}
\label{app:omega-proof}

The state calibration factor \(\omega_j\) rescales the raw Taylor fairness gain
according to the current income state, following the common optimization
principle of separating a candidate direction from its scale
\cite{nocedal2006numerical,conn2000trust}. Let
\(a=\sqrt{\mathcal{U}_{j-1}}\ge 0\) and \(b=\sqrt{\Lambda_j}\ge 0\), where
\(a\) measures the residual income-unfairness scale and \(b\) measures the local
second-order curvature scale. We define:
\begin{equation}
\small
\omega_j
=
\frac{1}{2}
+
\frac{a}{a+b+\epsilon},
\qquad \epsilon>0.
\end{equation}
When \(a\) and \(b\) are comparable, \(\omega_j\) is close to \(1\); when
\(a\) dominates, \(\omega_j>1\); and when \(b\) dominates, \(\omega_j<1\).
Thus, \(\omega_j\) strengthens the fairness correction when residual
unfairness is large and weakens it when the local curvature scale dominates.

Since \(\epsilon>0\),
\begin{equation}
\small
0\le \frac{a}{a+b+\epsilon}<1,
\qquad
\omega_j\in[1/2,3/2).
\end{equation}
Moreover,
\begin{equation}
\small
\frac{\partial \omega_j}{\partial a}
=
\frac{b+\epsilon}{(a+b+\epsilon)^2}>0,
\qquad
\frac{\partial \omega_j}{\partial b}
=
-\frac{a}{(a+b+\epsilon)^2}\le 0.
\end{equation}
Therefore, \(\omega_j\) increases with residual unfairness and decreases when
the local curvature scale dominates. Finally, since \(\omega_j>0\), for any
candidate \(d\),
\begin{equation}
\small
\operatorname{sign}\!\left(\omega_j G^{\mathrm{fair}}_j(d)\right)
=
\operatorname{sign}\!\left(G^{\mathrm{fair}}_j(d)\right).
\end{equation}
Thus, \(\omega_j\) is bounded, monotone, and sign-preserving. It adaptively
rescales the raw Taylor fairness gain without reversing its fairness direction.
\section*{GenAI Usage Disclosure}
The authors used GPT-5.5 only for language polishing and grammar checking. All technical ideas, algorithms, experiments, analyses, and writing decisions were made and verified by the authors.

% \section{Proof of State Calibration Properties}
% \label{app:omega-proof}

% Let \(a=\sqrt{\mathcal{U}_{j-1}}\ge 0\) and
% \(b=\sqrt{\Lambda_j}\ge 0\). Since \(\epsilon>0\),
% \[
% 0\le \frac{a}{a+b+\epsilon}<1,
% \]
% which directly gives \(\omega_j\in[1/2,3/2)\). Moreover,
% \[
% \frac{\partial \omega_j}{\partial a}
% =
% \frac{b+\epsilon}{(a+b+\epsilon)^2}>0,
% \qquad
% \frac{\partial \omega_j}{\partial b}
% =
% -\frac{a}{(a+b+\epsilon)^2}\le 0.
% \]
% Thus, the calibration increases with residual unfairness and decreases
% when the local curvature scale dominates. Finally, since
% \(\omega_j>0\), for every candidate \(d\),
% \[
% \operatorname{sign}\!\left(\omega_j G^{\mathrm{fair}}_j(d)\right)
% =
% \operatorname{sign}\!\left(G^{\mathrm{fair}}_j(d)\right).
% \]
% Therefore, \(\omega_j\) only applies a bounded positive rescaling to the
% raw Taylor fairness gain and cannot reverse its fairness direction.

\bibliographystyle{ACM-Reference-Format}
\bibliography{sample-base}

\end{document}